\pdfoutput=1
\documentclass[aps,prd,amsmath,floats,floatfix, twocolumn,
superscriptaddress,nofootinbib,showpacs,longbibliography]{revtex4-1}
\usepackage[T1]{fontenc}
\usepackage[utf8]{inputenc}
\usepackage{lmodern}
\usepackage{verbatim}
\usepackage[dvipsnames, usenames]{xcolor}
\definecolor{linkcolor}{rgb}{0.0,0.3,0.5}
\usepackage[hypertexnames=false, unicode, colorlinks=true, linkcolor=linkcolor,
citecolor=linkcolor, filecolor=linkcolor,urlcolor=linkcolor,
pdfusetitle]{hyperref}
\usepackage[all]{hypcap}
\usepackage{graphicx}
\usepackage{xspace}
\usepackage{amssymb}
\usepackage{microtype}

\usepackage{url}

\usepackage[english]{babel}
\usepackage{blindtext}

\usepackage[normalem]{ulem} 
\usepackage{bm} 
\usepackage{mathrsfs}

\usepackage[caption=false]{subfig}

\DeclareMathAlphabet{\mathpzc}{OT1}{pzc}{m}{it}

\usepackage[nomessages]{fp}

\begin{document}
\title{Study of eccentric binary black hole mergers using numerical relativity and an inspiral-merger-ringdown model}
\newcommand{\KITP}{\affiliation{Kavli Institute for Theoretical Physics,\\University of California Santa Barbara, Kohn Hall, Lagoon Rd, Santa Barbara, CA 93106}}

\author{Tousif Islam}
\email{tislam@kitp.ucsb.edu}
\KITP

\hypersetup{pdfauthor={Islam et al.}}

\date{\today}

\begin{abstract}
We study the phenomenology of non-spinning eccentric binary black hole (BBH) mergers using numerical relativity (NR) waveforms and \texttt{EccentricIMR} waveform model, as presented in Ref.~\cite{Hinder:2017sxy} (Hinder, Kidder, and Pfeiffer, \href{arXiv:1709.02007}{arXiv:1709.02007}). This model is formulated by combining an eccentric inspiral, derived from a post-Newtonian (PN) approximation including 3PN conservative and 2PN reactive contributions to the BBH dynamics, with a circular merger model. A distinctive feature of \texttt{EccentricIMR} is its two-parameter treatment, utilizing eccentricity and mean anomaly, to characterize eccentric waveforms. We implement the \texttt{EccentricIMR} model in \texttt{Python} to facilitate routine use. We then validate the model against 35 eccentric NR waveforms obtained from both the SXS and RIT NR catalogs. We find that \texttt{EccentricIMR} model reasonably match NR data for eccentricities up to $0.16$, specified at a dimensionless reference frequency of $x=0.07$, and mass ratios up to $q=4$. Additionally, we use this model as a tool for cross-comparing eccentric NR data obtained from the SXS and RIT catalogs. Furthermore, we explore the validity of a circular merger model often used in eccentric BBH merger modelling using both the NR data and \texttt{EccentricIMR} model. Finally, we use this model to explore the effect of mean anomaly in eccentric BBH mergers.
\end{abstract}

\maketitle

\section{Introduction}
Eccentric binary black hole (BBH) mergers are anticipated as one of the prominent sources of gravitational waves (GWs) in the current generation of detectors~\cite{Harry:2010zz,VIRGO:2014yos,KAGRA:2020tym}. However, most of the signals detected thus far by the LIGO-Virgo-KAGRA collaboration are consistent with quasi-circular mergers~\cite{LIGOScientific:2018mvr,LIGOScientific:2020ibl,LIGOScientific:2021usb,LIGOScientific:2021djp}. This is because the majority of binaries become sufficiently circularized as they enter the sensitivity band of current detectors. Nevertheless, it is expected that under specific conditions, such as if the binary is formed within globular clusters or galactic nuclei, they may retain significant eccentricity even as they enter the sensitivity band of the current generation detectors~\cite{Gondan:2020svr,Romero-Shaw:2019itr}. It is important to note that there are claims suggesting that certain events, such as GW190521~\cite{LIGOScientific:2020iuh}, might be produced by eccentric mergers with estimated eccentricities as high as $0.7$~\cite{RomeroShaw:2020thy,Gayathri:2020coq,CalderonBustillo:2020xms}. However, the lack of available faithful eccentric waveform models poses a significant challenge in efficiently characterizing eccentric signals, if indeed they exist.
Developing accurate and efficient waveform models for eccentric BBH mergers is therefore crucial. This endeavor typically involves pushing post-Newtonian (PN) approximations to higher orders, conducting high-accuracy numerical relativity (NR) simulations of eccentric BBH mergers, and subsequently constructing semi-analytical or data-driven models utilizing the existing and forthcoming NR data. 

Indeed, initial efforts are underway to construct eccentric waveform models.
Refs.~\cite{Tiwari:2019jtz, Huerta:2014eca, Moore:2016qxz, Damour:2004bz, Konigsdorffer:2006zt, Memmesheimer:2004cv} have introduced ready-to-use waveform models, in both time and frequency domains, for binaries moving in inspiralling eccentric orbits. These models are constructed based on purely post-Newtonian (PN) approximations up to varying orders.
Refs.~\cite{Hinder:2017sxy, Cho:2021oai} have taken a step further by combining a PN eccentric inspiral model with a quasi-circular merger model, thereby providing a comprehensive eccentric inspiral-merger-ringdown model. These models assume that binaries with initially moderate eccentricities tend to sufficiently circularize by the time they transition from inspiral to the merger stage. A slightly different approach is presented in Ref.~\cite{Chattaraj:2022tay} that first hybridizes eccentric PN inspiral with NR data and then builds an analytical representation of the waveform. All these models however only include the quadrupolar mode of radiation and ignore all higher order spherical harmonic modes.

On the other hand, Refs.~\cite{Hinderer:2017jcs,Cao:2017ndf,Chiaramello:2020ehz,Albanesi:2023bgi,Albanesi:2022xge,Riemenschneider:2021ppj,Chiaramello:2020ehz,Ramos-Buades:2021adz,Liu:2023ldr} have presented eccentric waveform models within the effective-one-body (EOB) formalism, calibrating them against a handful of available eccentric NR simulations. These models typically extend non-eccentric NR-calibrated EOB models by incorporating eccentric corrections up to certain PN orders. The treatment of the merger-ringdown stage usually assumes a quasi-circular merger.
Another class of models employs a consistent combination of post-Newtonian (PN) approximation, self-force, and black-hole perturbation theory to describe the eccentric inspiral, followed by a quasi-circular merger model~\cite{Huerta:2016rwp,Huerta:2017kez,Joshi:2022ocr}. There are also recent efforts to use NR data directly to establish a mapping between quasi-circular and eccentric waveforms in the time-domain~\cite{Setyawati:2021gom,Wang:2023ueg}. These studies aim to understand the eccentricity-induced modulations in amplitudes and frequencies and develop a semi-analytical description of these phenomena. While some of these models include higher order modes, they are not very accurate.

A distinct approach was proposed in Ref.~\cite{Islam:2021mha}, which explored data-driven strategies to construct an eccentric IMR model using NR data without assuming a quasi-circular merger. The study presents a proof-of-principle eccentric model, \texttt{NRSur2dq1Ecc} for equal mass BBH systems with eccentricities of up to $0.2$, as measured about 20 cycles before the merger. The model is found to be applicable up to a mass ratio of $q:=m_1/m_2=3$ (where $m_1$ and $m_2$ are the masses of the primary and secondary black holes, respectively) for small eccentricities.

Among all these models, only the ones presented in Refs.\cite{Hinder:2017sxy,Cho:2021oai,Islam:2021mha,Ramos-Buades:2021adz} use a two-parameter treatment to characterize an eccentric waveform. These models incorporate both eccentricity and the mean anomaly as parameters to describe an eccentric BBH evolution. This dual-parameter approach makes them inherently more suitable for data analysis than models that either fix the mean anomaly parameter or do not consider it at all. However, the \texttt{NRSur2dq1Ecc} model is primarily valid for equal mass binaries, and the model presented in Refs.\cite{Cho:2021oai} has not been tested against NR simulations yet. Eccentric EOB model developed in Ref.~\cite{Ramos-Buades:2021adz} is not currently available publicly. On the other hand, the \texttt{EccentricIMR} model, presented in Ref.~\cite{Hinder:2017sxy}, has been validated against 23 eccentric SXS NR simulations, demonstrating reasonable accuracy and available for public use. 

We therefore find \texttt{EccentricIMR} model suitable to use in understanding the phenomenology of the eccentric BBH mergers and to use it in future multi-query source characterization studies. As this model uses PN to provide inspiral waveforms, using \texttt{EccentricIMR} will also help in understanding validity of PN approximations in eccentric binary mergers. We first revisit the of \texttt{EccentricIMR} model against NR data. The original validation method in Ref.~\cite{Hinder:2017sxy} involved comparing the eccentric PN inspiral to NR data from the SXS collaboration, very close to the merger, which may not be optimal due to potential limitations of the post-Newtonian approximation in that regime. To address this, we employ an alternative validation scheme and re-evaluates the model against a set of 15 SXS NR data. Additionally, the model's accuracy are tested against a new set of 23 eccentric NR simulations from the RIT catalog, covering mass ratios up to $q=7$. Using these new high mas ratio eccentric NR simulations, we then examine the validity of a circular merger model used in several earlier eccentric waveform models. We further use \texttt{EccentricIMR} model as an intermediate step to cross-compare SXS and RIT NR data. Finally, we study eccentric BBH waveform phenomenology, specially the effect of mean anomaly parameter, using \texttt{EccentricIMR} model.

The rest of the paper is organized as follows. In Section~\ref{sec:eccentricimr}, we provide an executive summary of the \texttt{EccentricIMR} model. Subsequently, in Section~\ref{sec:nr_validation}, we discuss NR simulations obtained from SXS and RIT catalogs and validate the \texttt{EccentricIMR} model against them. In Section~\ref{sec:cmm}, we delve into the merger-ringdown structure of eccentric NR waveforms and examine the validity of a circular merger approximation. Section~\ref{sec:phenomenology} then explores the phenomenology of eccentric BBH waveforms. Finally, in Section~\ref{sec:discussion}, we discuss the implications of our results and outline future plans.

\section{Eccentric waveform model}
\label{sec:eccentricimr}
Gravitational radiation (waveform) from a BBH merger is written as a superposition of $-2$ spin-weighted spherical harmonic modes with indices $(\ell,m$):
\begin{align}
h(t,\theta,\phi;\boldsymbol{\lambda}) &= \sum_{\ell=2}^\infty \sum_{m=-\ell}^{\ell} h^{\ell m}(t;\boldsymbol\lambda) \; _{-2}Y_{\ell m}(\theta,\phi)\,,
\label{hmodes}
\end{align}
where $\boldsymbol{\lambda}$ is the set of intrinsic parameters (such as the masses and spins of the binary) describing the binary, and ($\theta$,$\phi$) are angles describing the orientation of the binary.
Each spherical harmonic mode $h_{\ell m}(t)$ is a complex time series and is further decomposed into a real amplitude $A_{\ell m}(t)$ and phase $\phi_{\ell m}(t)$, as
\begin{equation}
h_{\ell m}(t) = A_{\ell m}(t) e^{i \phi_{\ell m}(t)} \,.
\label{eq:amp_phase}
\end{equation}
Instantaneous frequency of each spherical harmonic mode is given by:
\begin{equation}
\omega_{\ell m}(t) = \frac{d\phi_{\ell m}(t)}{dt} \, .
\label{eq:freq}
\end{equation}
We choose the time axis such a way that $t=0$ denote the maximum amplitude of the $(\ell,m)=(2,2)$ spherical harmonic mode. 

To generate eccentric waveforms, we utilize the \texttt{EccentricIMR} model presented at \href{https://github.com/ianhinder/EccentricIMR}{https://github.com/ianhinder/EccentricIMR}. This model uses post-Newtonian inspiral waveform and their quasi-circular merger model (CMM) to obtain a time-domain eccentric IMR waveform. Inspiral part of the waveform includes 3PN order conservative and 2PN order reactive contributions to the BBH dynamics~\cite{Hinder:2008kv}. While the original model is written in \texttt{Mathematica}, for ease of use, we create a Python wrapper on top of it. The model can generate gravitational waveforms for non-spinning eccentric BBH mergers and takes mass ratio $q$, initial eccentricity $e_0$, initial mean anomaly $l_0$, and a starting dimensionless frequency $x_0=(M \omega_{\text{orb,0}})^{2/3}$ as inputs. Here, $\omega_{\text{orb}}$ is the orbital angular frequency of the binary and can be obtain as: 
\begin{equation}
    \omega_{\text{orb}}=0.5 \times \omega_{22}.
\end{equation}

\begin{figure}
\includegraphics[width=\columnwidth]{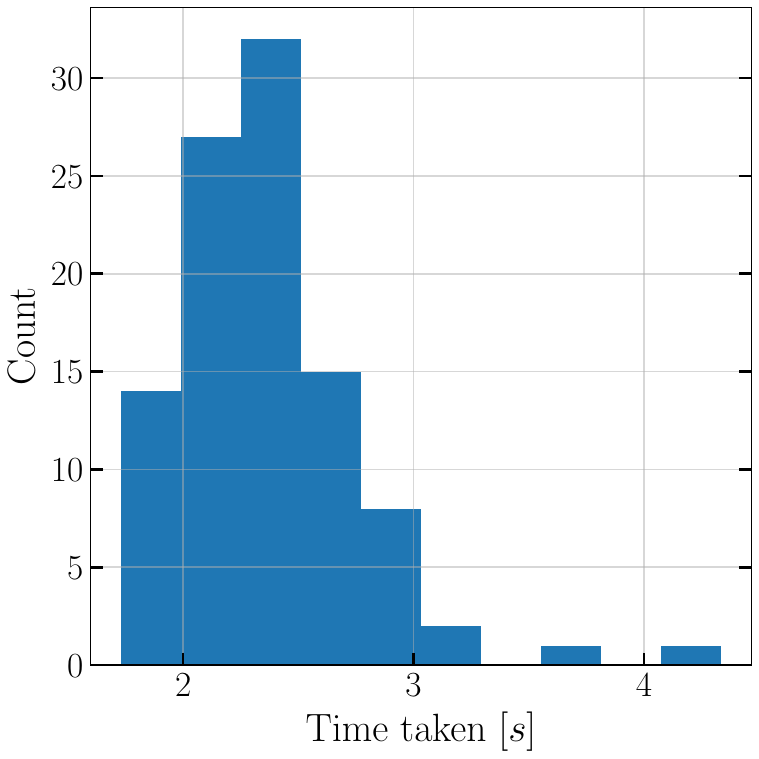}
\caption{Distribution of the time taken to generate an eccentric waveform using the Python wrapper of the \texttt{EccentricIMR} model. More details are in Section~\ref{sec:eccentricimr}.}
\label{fig:timing}
\end{figure}

We generate a total of 100 random eccentric waveforms with mass ratios ranging from $q=1$ to $q=4$, eccentricities $e_0=0.0$ to $e_0=0.15$, and mean anomaly $l_0=-\pi$ to $l_0=\pi$ as given at $x_0=0.07$. In Figure~\ref{fig:timing}, we show the distribution of the time taken to generate an eccentric waveform using the Python wrapper of the \texttt{EccentricIMR} model. We find that typical waveform generation takes $\sim 2.5s$. This timing exercise reveals that while this model might not be suitable for real-time data analysis, it serves as an excellent tool for understanding eccentric BBH waveforms, particularly due to its use of two-parameter ($e_0$ and $l_0$) eccentricity definitions.

\section{Numerical relativity validation}
\label{sec:nr_validation}
Before we proceed to the NR validation of the \texttt{EccentricIMR} model, it is essential to note that the model has already undergone testing against a set of 23 SXS NR simulations with mass ratios ranging from $q=1$ to $q=3$ and eccentricities $e\leq 0.07$, as measured about 7 cycles before the merger~\cite{Hinderer:2017jcs}. Here, we validate the model against both the SXS NR data and newly available RIT NR data. While the validation against SXS NR, using roughly the same set of NR simulations, serves as a sanity check of our Python implementation, the validation against RIT NR data serves two purposes. First, it offers more testing for the \texttt{EccentricIMR} model. Second, it helps in characterizing the RIT NR simulations in a more systematic way.

\begin{figure}
\includegraphics[width=\columnwidth]{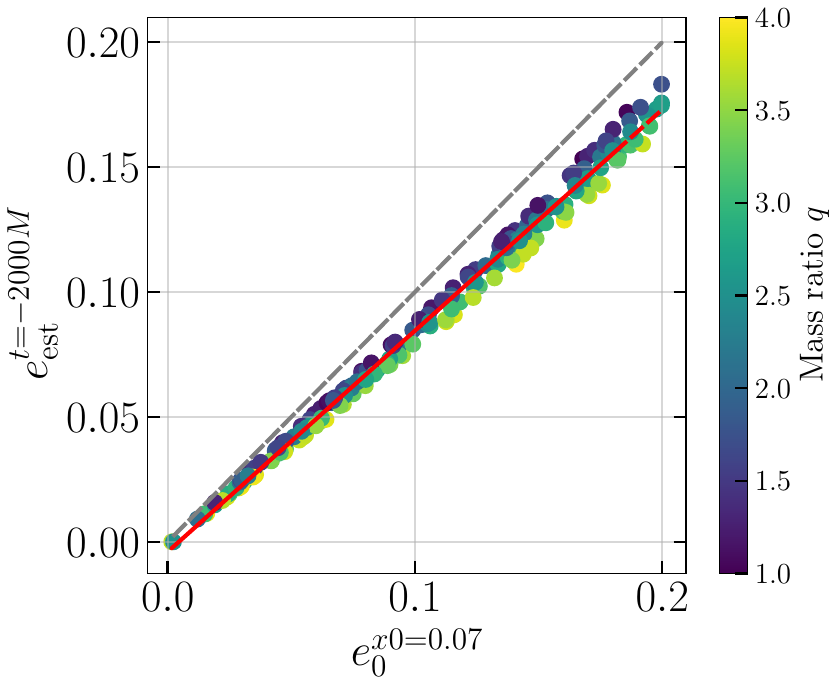}
\caption{We show the PN eccentricity $e_{0}^{x_0=0.07}$, estimated at the reference frequency of $x_0=0.07$, and the eccentricity estimator $e_{\rm est}^{t=-2000M}$, calculated using Eq.(\ref{eq:ecc}) at a reference time of $t=-2000M$. We color-code the points by their respective mass ratios. The dashed grey line indicates $e_{0}^{x_0=0.07}=e_{\rm est}^{t=-2000M}$. The solid red line represents an overall linear fit to the data. More details are in Section~\ref{sec:mapping}.}
\label{fig:e0_eIMR}
\end{figure}

\subsection{NR data}
We utilize a total of 15 publicly available SXS eccentric NR simulations with mass ratios ranging from $q=1$ to $q=3$ and eccentricities $e\leq 0.07$, as measured approximately 7 cycles before the merger. We exclude 5 simulations from our study due to their high eccentricity~\footnote{The Python implementation encounters difficulties related to Wolfram kernel while generating waveforms with high eccentricity}. Additionally, we incorporate a total of 20 eccentric simulations from the RIT catalog to assess the performance of the \texttt{EccentricIMR} model. These RIT simulations exhibit diverse durations, ranging from $\sim 1000M$ to $\sim 8000M$, with eccentricities reaching up to $\sim 0.2$, and mass ratios spanning between $q=1$ and $q=4$. Furthermore, we use an additional 3 RIT eccentric NR data with mass ratios up to $q=7$ to evaluate the validity of a circular merger model.

\begin{figure*}
\includegraphics[width=\textwidth]{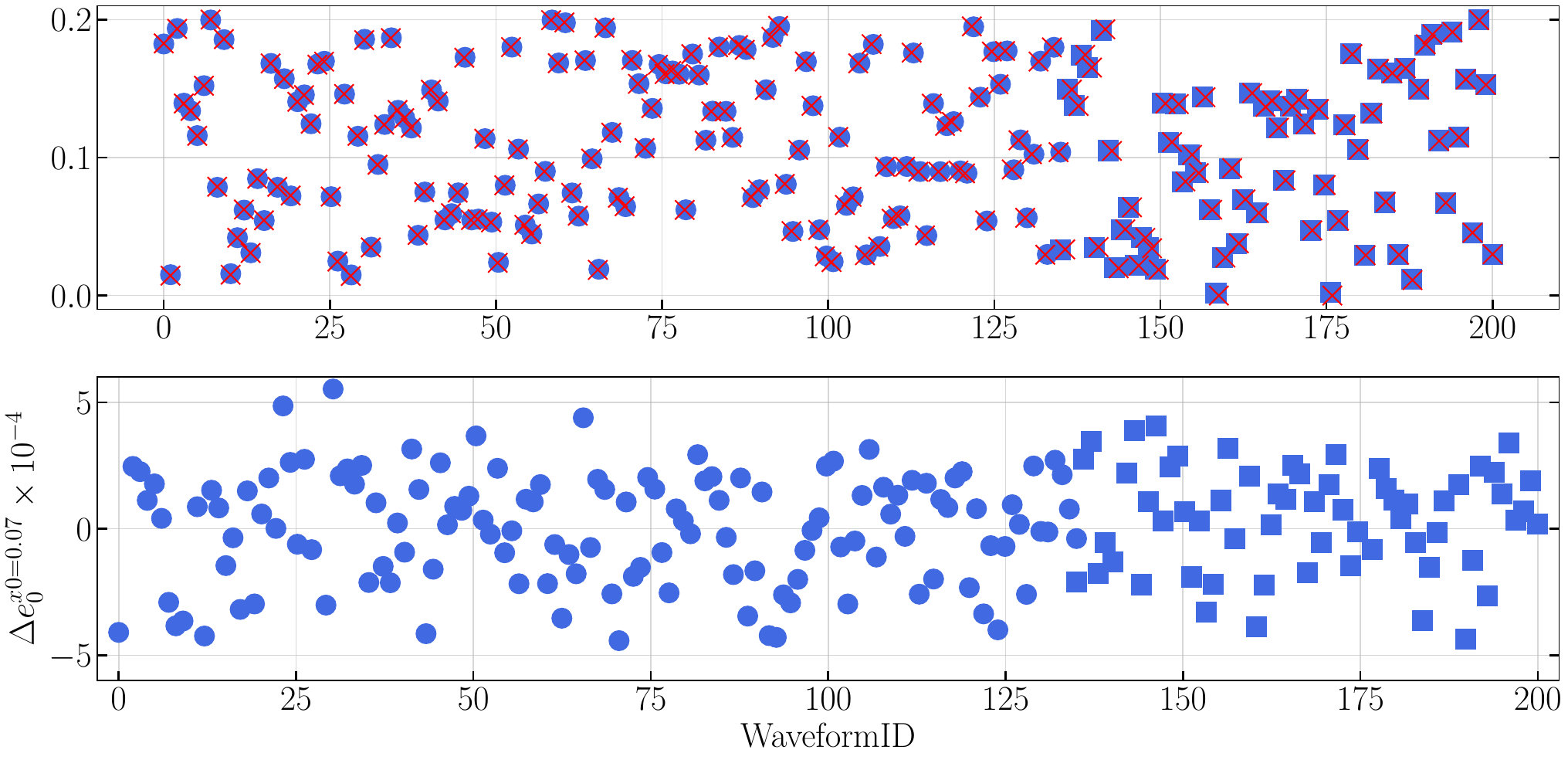}
\caption{We demonstrate the fitting accuracy of our eccentricity mapping between $e_{0}^{x_0=0.07}$ and $e_{\rm est}^{t=-2000M}$. Blue circles indicate training points, and blue squares indicate validation points. The upper panel shows the exact eccentricity values, while the lower panel shows the differences between exact values and fit predictions. More details are in Section~\ref{sec:mapping}.}
\label{fig:eccentricity_analytical_fit}
\end{figure*}

\subsection{Comparing \texttt{EccentricIMR} model to NR}
Below, we provide an executive summary of the tools and methods used in comparing the \texttt{EccentricIMR} model to NR. In particular, it involves estimating eccentricities from the waveforms, developing a mapping between PN eccentricity and eccentricity estimators, and finally optimizing PN eccentricity parameters so that they yield NR-faithful waveforms.

\subsubsection{Common eccentricity estimators} 
We characterize eccentric BBH signals using two common eccentricity parameters: eccentricity $e_{\rm est}$ and mean anomaly $l_{\rm est}$. Both of these quantities are directly computed from the waveform and can thus be applied to any eccentric waveform, regardless of its generation method. Our eccentricity estimator is defined as~\cite{Mora:2002gf}:
\begin{equation} \label{eq:ecc}
e_{\rm est}(t) = \frac{\sqrt{\omega_{\text{orb},p}(t)}-\sqrt{\omega_{\text{orb},a}(t)}}{\sqrt{\omega_{\text{orb},p}(t)}+\sqrt{\omega_{\text{orb},a}(t)}},
\end{equation}
where $\omega_{\text{orb},p}(t)$ and $\omega_{\text{orb},a}(t)$ represent the orbital frequencies at each periastron (the point where the black holes are at their closest distance) and apastron (the point where the black holes are at their farthest distance from each other).
Note that $\omega_{\text{orb},p}$ and $\omega_{\text{orb},a}$ are discrete points in time. However, we identify all discrete $\omega_{\text{orb},p}$ and $\omega_{\text{orb},a}$ values, along with their respective time coordinates, to construct continuous spline representations. These splines are then employed to obtain a continuous representation of the eccentricity $e_{\rm est}(t)$. 

On the other hand, the mean anomaly estimator is simply
\begin{equation}\label{eq:ano}
l_{\rm est}(t) = \frac{2\pi (t-t_0)}{P},
\end{equation}
where $P$ is the period of the respective orbit, and $t_0$ is the time of the immediate last periastron passage. We compute $P$ by calculating the time difference between two consecutive periastrons.

\subsubsection{Mapping between eccentricity estimators and PN eccentricity} 
\label{sec:mapping}
Next, we establish a mapping between the eccentricity estimators $\{e_{\rm est}, l_{\rm est}\}$, obtained at a reference time of $t=-2000M$, and the PN eccentricity parameters $\{e_0, l_0\}$, given at a dimensionless reference frequency of $x_0=0.07$. To achieve this, we randomly generate 200 waveforms using the \texttt{EccentricIMR} model for binaries with mass ratios ranging from $q=1$ to $q=4$ and eccentricities ranging from $e_0=0.0$ to $e_0=0.15$ and mean anomaly ranging from $l_0=-\pi$ to $l_0=\pi$. Subsequently, we estimate $\{e_{\rm est}^{t=-2000M}, l_{\rm est}^{t=-2000M}\}$ at $t=-2000M$ for each waveform.

\begin{figure}
\includegraphics[width=\columnwidth]{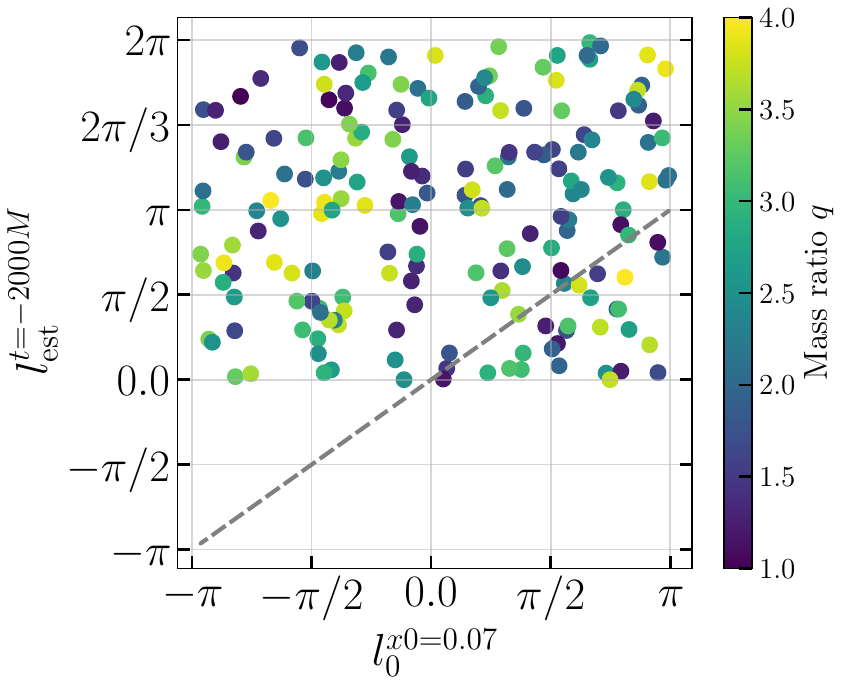}
\caption{We show the PN mean anomaly $l_{0}^{x_0=0.07}$, given at the reference frequency of $x_0=0.07$, and the mean anomaly estimator $l_{\rm est}^{t=-2000M}$, calculated using Eq.(\ref{eq:ecc}) at a reference time of $t=-2000M$. We color-code the points by their respective mass ratios. The dashed grey line indicates $l_{0}^{x_0=0.07}=l_{\rm est}^{t=-2000M}$. More details are in Section~\ref{sec:mapping}.}
\label{fig:l0_lIMR}
\end{figure}

In Figure~\ref{fig:e0_eIMR}, we show the eccentricity estimator $e_{\rm est}^{t=-2000M}$ as a function of the initial PN eccentricity $e_0^{x_0=0.07}$. The values are color-coded based on the mass ratios. We observe that $e_{\rm est}^{t=-2000M}$ is consistently smaller than the initial eccentricity values $e_0^{x_0=0.07}$. This is not surprising as $t=-2000M$ corresponds to a later time in the binary evolution and therefore eccentricity values decrease. Additionally, we notice an almost linear relationship between these two eccentricity parameters for any given mass ratios. Therefore, we perform an overall linear fit of the initial eccentricity values and obtain:
\begin{equation}
    e_{0}^{x_0=0.07} = 0.8850844 \times e_{\rm est}^{t=-2000M}  - 0.00395882.
\end{equation}
While this fit provides a rough prediction of the eccentricity estimator at a reference time of $t=-2000M$, it lacks a clear mass ratio dependence. Therefore, we generalize the fit and obtain:        
\begin{align}
    e_{0}^{x_0=0.07}(q,e_{\rm est}^{t=-2000M}) = f_1(q) \times f_2(e_{\rm est}^{t=-2000M})
\end{align}
with
\begin{align}
    f_1(q) = & 5.607 \times 10^{-1} \times q - 8.961 \times 10^{-1} \times q^2 \notag \\ 
    &-9.763 \times 10^{-2} \times q^3 -4.382 \notag \\
\end{align}
and
\begin{align}
    f_2(e_{\rm est}^{t=-2000M}) = &-2.697 \times 10^{_2} e_{\rm est}^{t=-2000M} \notag \\
    &+1.028 \times 10^{-2} \times (e_{\rm est}^{t=-2000M})^2 \notag \\
    &+ 2.659 \times 10^{-2} \times (e_{\rm est}^{t=-2000M})^3  \\
\end{align}
To construct the fit, we divide our data points into two categories. We randomly assign 135 points for training and the remaining 65 points for validation. We demonstrate the fitting accuracy in Figure~\ref{fig:eccentricity_analytical_fit}. We find that our analytical fit can predict $e_{\rm est}^{t=-2000M}$ with a maximum error of $\sim 0.0005$.

Figure~\ref{fig:l0_lIMR} then shows the mean anomaly estimator $l_{\rm est}^{t=-2000M}$ as a function of the initial PN mean anomaly $l_0^{x_0=0.07}$. The values are color-coded based on the mass ratios. The first notable observation is the range of mean anomaly values. While $l_0^{x_0=0.07}$ ranges from $-\pi$ to $\pi$, the range of $l_{\rm est}^{t=-2000M}$ is $[0,2\pi]$. We also notice that unlike in the case of eccentricity, here we do not find a clear trend between $l_0^{x_0=0.07}$ and $l_{\rm est}^{t=-2000M}$. Our efforts to build meaningful fits for $l_0^{x_0=0.07}$ also remains unsuccessful.

\begin{figure}
\includegraphics[width=\columnwidth]{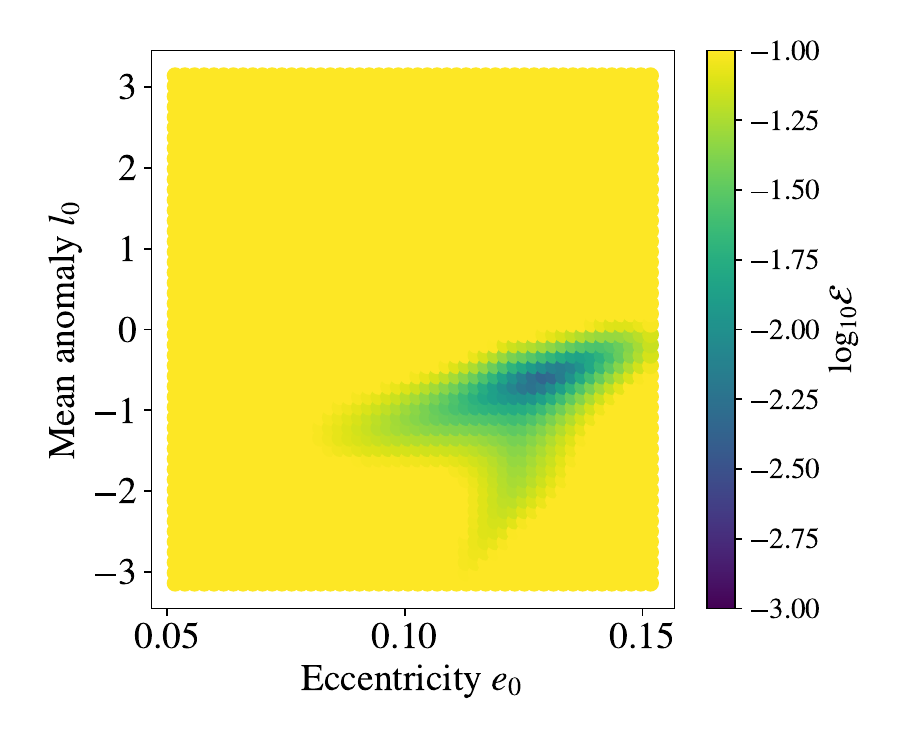}
\caption{We show the time-domain errors $\mathcal{E}$, between \texttt{EccentricIMR} waveforms and NR data obtained from the \texttt{SXS:BBH:1359} simulation, as a function of the eccentricity $e_0$ and mean anomaly $l_0$. We use reference dimensionless frequency $x_0=0.07$ and mass ratio $q=1$. More details are in Section~\ref{sec:sxsnr}.}
\label{fig:sxs1359_grid}
\end{figure}

\subsubsection{Matching \texttt{EccentricIMR} waveforms to NR} 
To compare \texttt{EccentricIMR} waveforms to NR, we need to ensure that we are generating waveforms for the same system. One approach is to ensure they start from the same eccentricity values at a common reference time or frequency. However, this becomes complicated as the eccentricity for the NR data is typically quoted at the start of the NR simulations, while \texttt{EccentricIMR} waveforms are characterized by the eccentricities at the beginning of the PN inspiral. Duration of the NR simulations vary widely. Furthermore, depending on the input start frequencies, \texttt{EccentricIMR} waveforms also have different lengths. 

\begin{figure*}
\includegraphics[width=\textwidth]{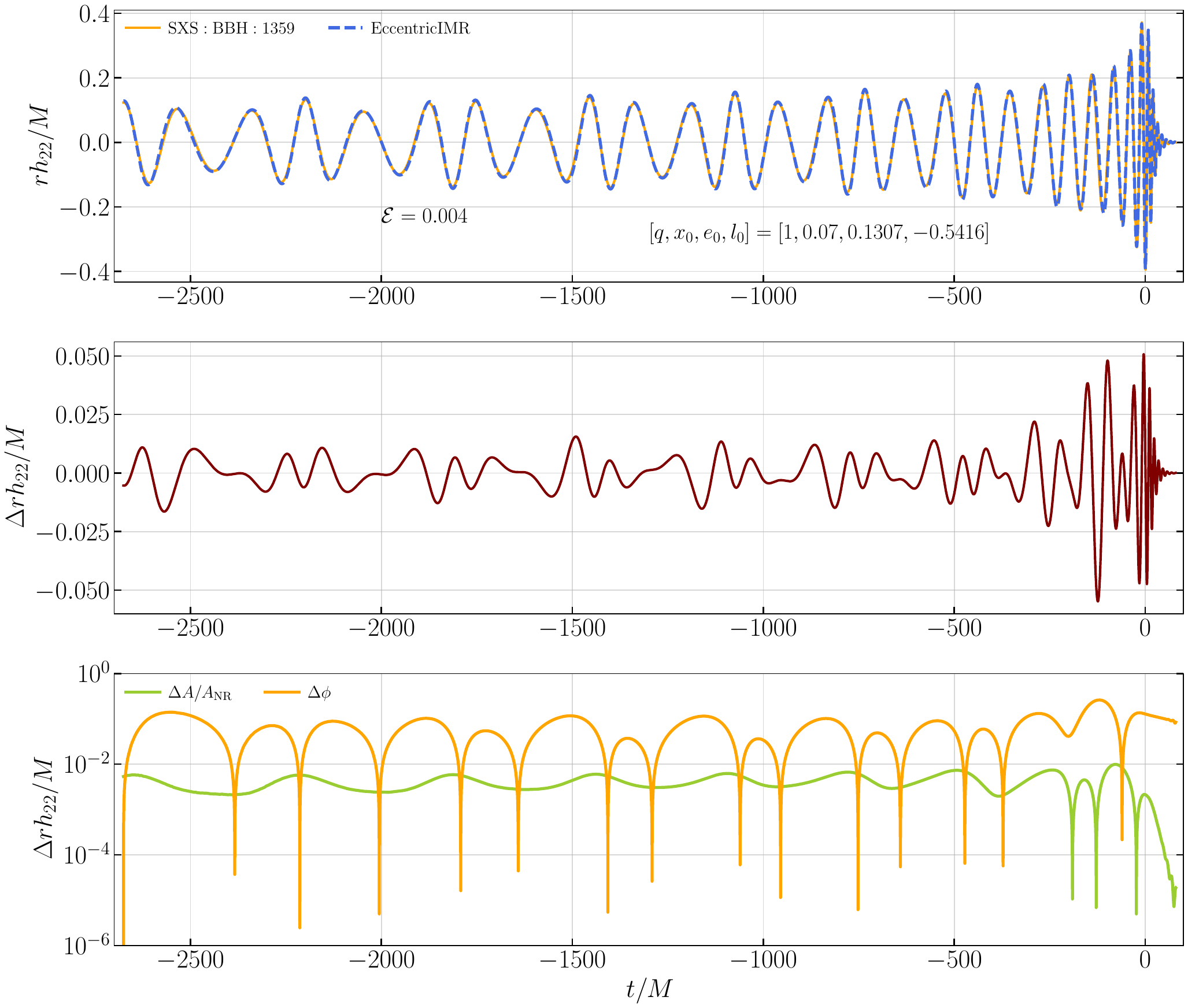}
\caption{We show the NR data obtained from the \texttt{SXS:BBH:1359} simulation (blue dashed line) and the optimized \texttt{EccentricIMR} waveform (orange solid line) generated with $[q,x_0,e_0,l_0]=[1,0.07,0.1307,-0.5416]$ in the upper panel. Additionally, we show the difference between these two waveforms in the middle panel and the differences in amplitude and phase in the lower panel. More details are in Section~\ref{sec:sxsnr}.}
\label{fig:sxs1359}
\end{figure*}

We, therefore, employ a slightly different approach. For each NR data, we estimate the eccentricity $e_{\rm est}^{t=-2000M}$ and mean anomaly $l_{\rm est}^{t=-2000M}$ at a reference time of $t=-2000M$. Subsequently, we use the analytical mapping developed in Section~\ref{sec:mapping} to obtain a crude initial guess for the initial eccentricity $e_{0}^{x_0=0.07}$ required in the \texttt{EccentricIMR} model to match NR. For the mean anomaly, our initial guess value is set to $l_{\rm est}^{t=-2000M}$. We then initiate an optimization process with these initial guesses to determine the values for $\{e_0^{x_0=0.07},l_0^{x_0=0.07}\}$ that minimize the time-domain error between NR and \texttt{EccentricIMR} waveforms after time/phase alignment:
\begin{equation}
\mathcal{E} = \int_{t_{\rm ini}}^{t_{\rm final}} \frac{|h_{\tt NR}(q; t) - h_{\tt IMR}(q,e_{0}^{x_0},l_{0}^{x_0}; t)|^2}{|h_{\tt NR}|^2} dt,
\end{equation}
where $t_{\rm ini}$ and $t_{\rm final}$ represent the initial and end times of the waveforms. Note that the mass ratio value $q$ is same between NR and \texttt{EccentricIMR} model. We cast both the waveforms on a common time grid and ensure that the peak amplitude is at $t=0$. We then align them such a way that the starting orbital phase is zero. To perform the optimization, we utilize the \texttt{scipy.optimize} module with the \texttt{nelder-mead} method.

While this method works efficiently for most of the NR data, some RIT NR simulations are significantly short in length, often covering only the last $1000M$ of the binary evolution. This implies that the waveform only exhibits three to four periastron or apastron passages. Consequently, attempts to estimate eccentricity or mean anomaly using the waveform fail, as the data is insufficient to build a faithful spline representation. Therefore, we need an alternative strategy to obtain initial guess for the optimization process.

For NR data with significantly shorter duration, we adopt a two-step process to estimate the initial eccentricity $e_{0}^{x_0=0.07}$ and mean anomaly $l_{0}^{x_0=0.07}$. Firstly, we create a two-dimensional uniform grid with $2500$ points for $e_{0}^{x_0=0.07}$ ranging from $0.0$ to $0.15$ and mean anomaly $l_{0}^{x_0=0.07}$ ranging from $-\pi$ to $\pi$. Subsequently, we generate \texttt{EccentricIMR} waveforms at each point and compute the time-domain error against the NR data. We then select the $\{e_{0}^{x_0=0.07}, l_{0}^{x_0=0.07}\}$ values that yield the smallest error and use them as our initial guess for the optimization process described in the previous paragraph. 

This way, we can obtain optimized \texttt{EccentricIMR} waveforms for each NR data efficiently irrespective of their lengths.

\begin{figure*}[htp]
	\subfloat[\texttt{SXS:BBH:1361}]{
		\includegraphics[scale=0.5]{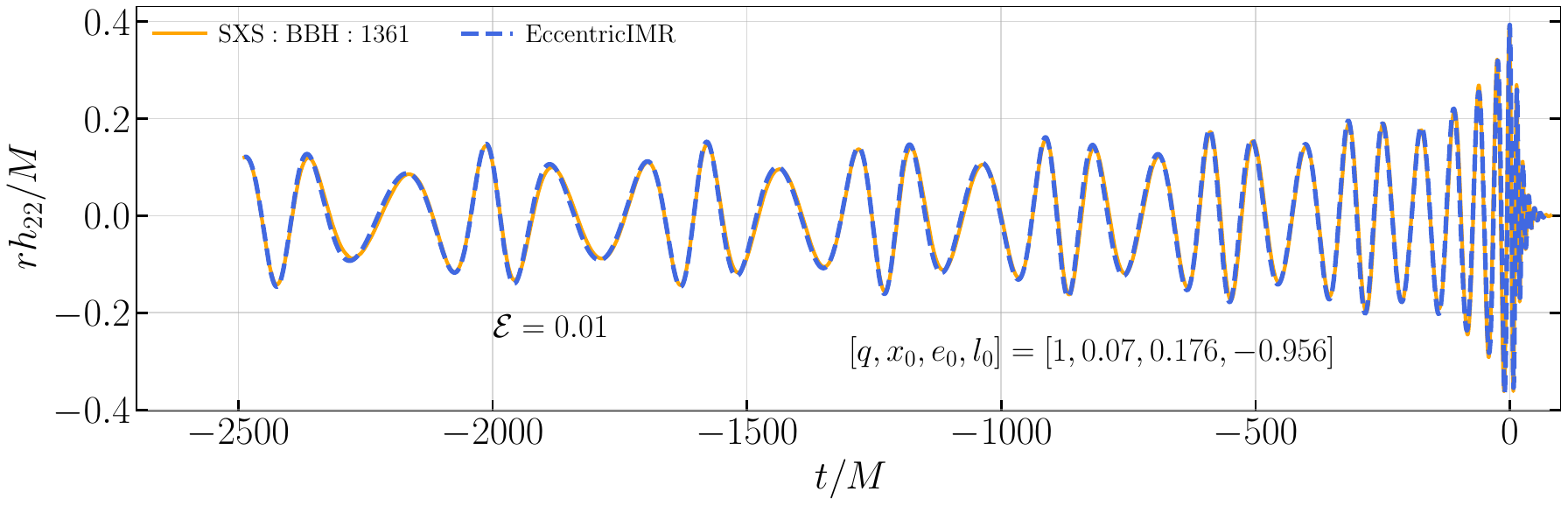}
		\label{fig:1361}
	}\\
	\subfloat[\texttt{SXS:BBH:1368}]{
		\includegraphics[scale=0.5]{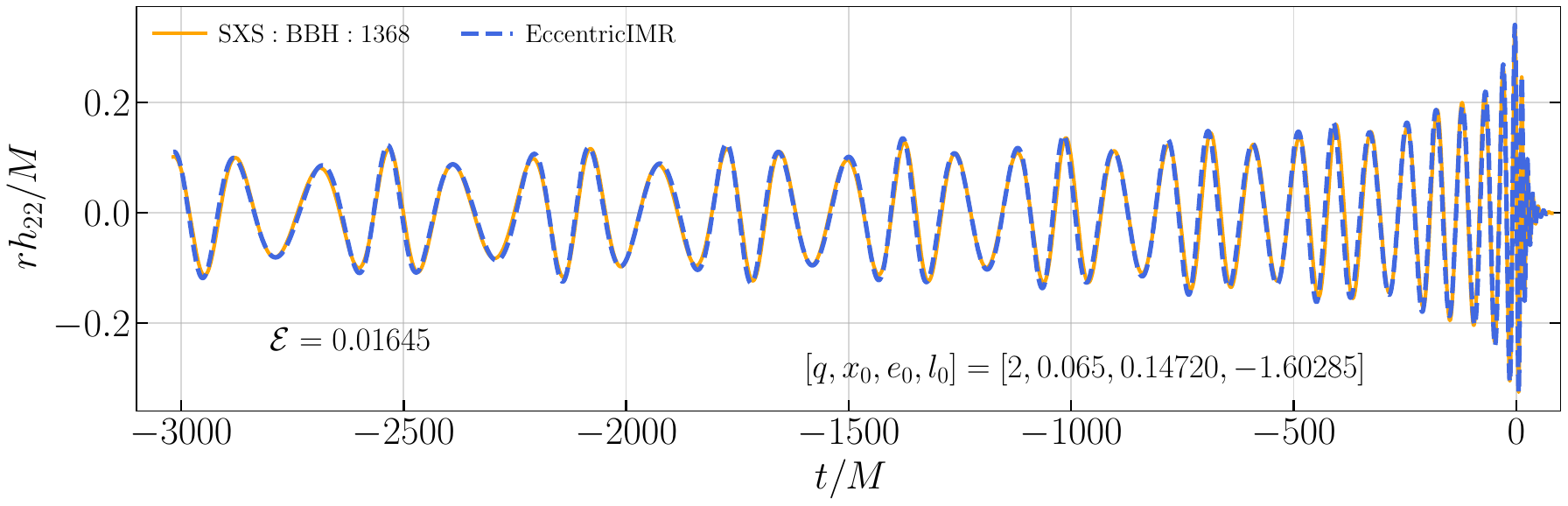}
		\label{fig:1368}
	}\\
    \subfloat[\texttt{SXS:BBH:1372}]{
		\includegraphics[scale=0.5]{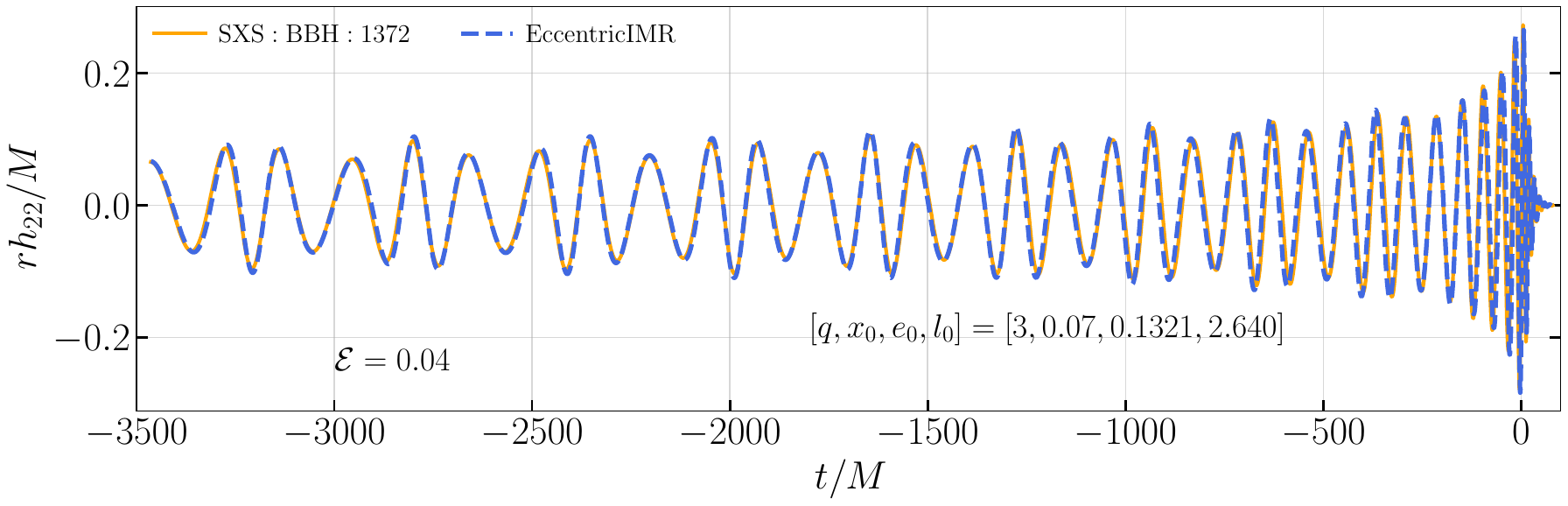}
		\label{fig:1372}
	}\\
	\subfloat[\texttt{SXS:BBH:1373}]{
		\includegraphics[scale=0.5]{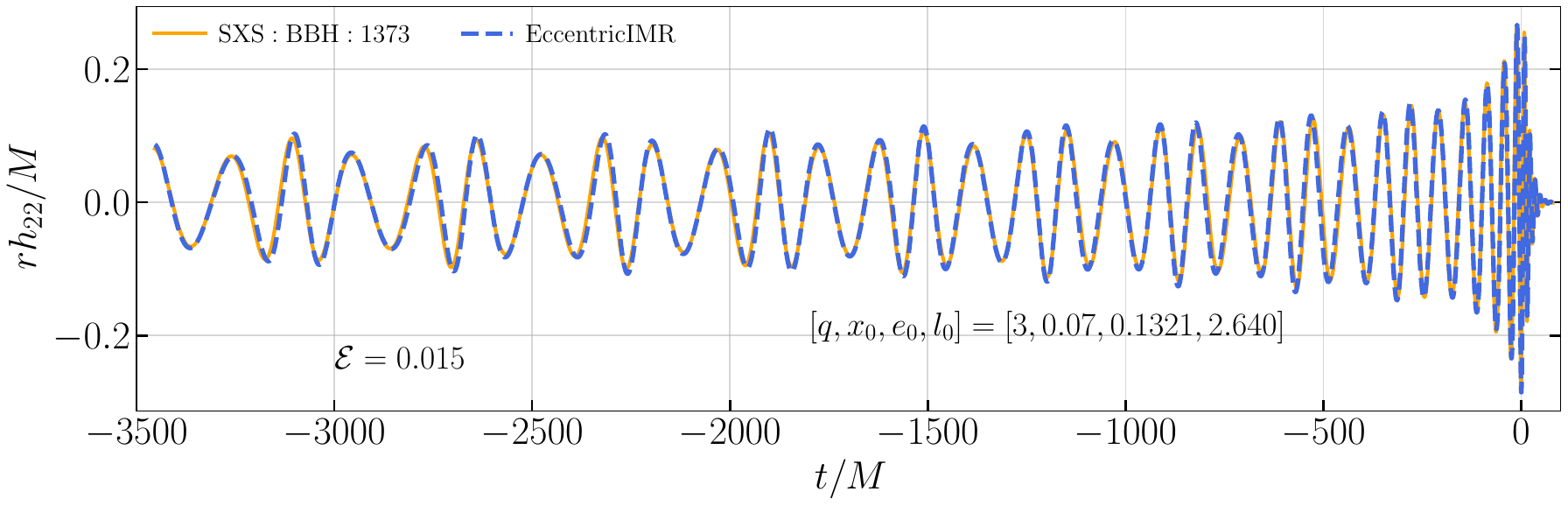}
		\label{fig:1373}
	}
\label{fig:SXSNR}
\caption{We show four representative highly eccentric SXS NR simulations (blue dashed lines), namely, (a) \texttt{SXS:BBH:1361}, (b) \texttt{SXS:BBH:1368}, (c) \texttt{SXS:BBH:1372} and (d) \texttt{SXS:BBH:1373}, alongside the corresponding optimized \texttt{EccentricIMR} waveforms (orange solid lines). These simulations have eccentricities ranging from $e_0=0.132$ to $e_0=0.176$, given at a reference frequency of $x_0=0.07$. More details are in Section~\ref{sec:sxsnr}.}
\end{figure*}

\subsection{Validation against SXS NR data}
\label{sec:sxsnr}
We first re-validate \texttt{EccentricIMR} waveforms against the SXS NR data. It is essential to note that our validation procedure differs slightly from the one employed in Ref~\cite{Hinder:2017sxy}. We conduct a global optimization to determine the best-fit values for $\{e_{\tt IMR}^{x_0=0.07}, l_{\tt IMR}^{x_0=0.07}\}$. In contrast, the Ref.~\cite{Hinder:2017sxy} matched the PN frequency to NR shortly before the merger to obtain eccentricity values at a reference frequency. Our validation experiment therefore would be sanity check that we can achieve similar level of accuracy reported in Ref.~\cite{Hinder:2017sxy}.

We demonstrate the effectiveness of our strategy and the faithfulness of the \texttt{EccentricIMR} model in Figure~\ref{fig:sxs1359}. We present the NR data obtained from the \texttt{SXS:BBH:1359} simulation alongside the optimized \texttt{EccentricIMR} waveform generated with $[q,x_0,e_0,l_0]=[1,0.07,0.1307,-0.5416]$. We observe that the \texttt{EccentricIMR} waveform closely matches with the NR data, showing no visual difference. We compute the time-domain error to be $0.004$, indicating a good match. Additionally, we analyze the difference between these two waveforms (Figure~\ref{fig:sxs1359}, middle panel). We notice that the difference exhibits clearly identifiable time-series features. This could indicate higher-order PN corrections. It is important to note that the \texttt{EccentricIMR} model includes up to 3PN terms in the inspiral. Furthermore, we show the differences in amplitude and phase in the lower panel of Figure~\ref{fig:sxs1359}. We find that the relative amplitude differences are $\sim 10^{-2}$, whereas the absolute phase differences are $\sim10^{-1}$. Finally, to demonstrate that the optimized values quoted corresponds to the minimum time-domain error, we show the time-domain errors as a function of the eccentricity $e_0$ and mean anomaly $l_0$ (we fix reference frequency $x_0=0.07$) in Figure~\ref{fig:sxs1359_grid}. We notice that even when eccentricity values are close to the optimized $e_0$, not all mean anomaly values yield reasonable match.

\begin{figure*}[htp]
	\subfloat[\texttt{RIT:eBBH:1282}]{
		\includegraphics[scale=0.5]{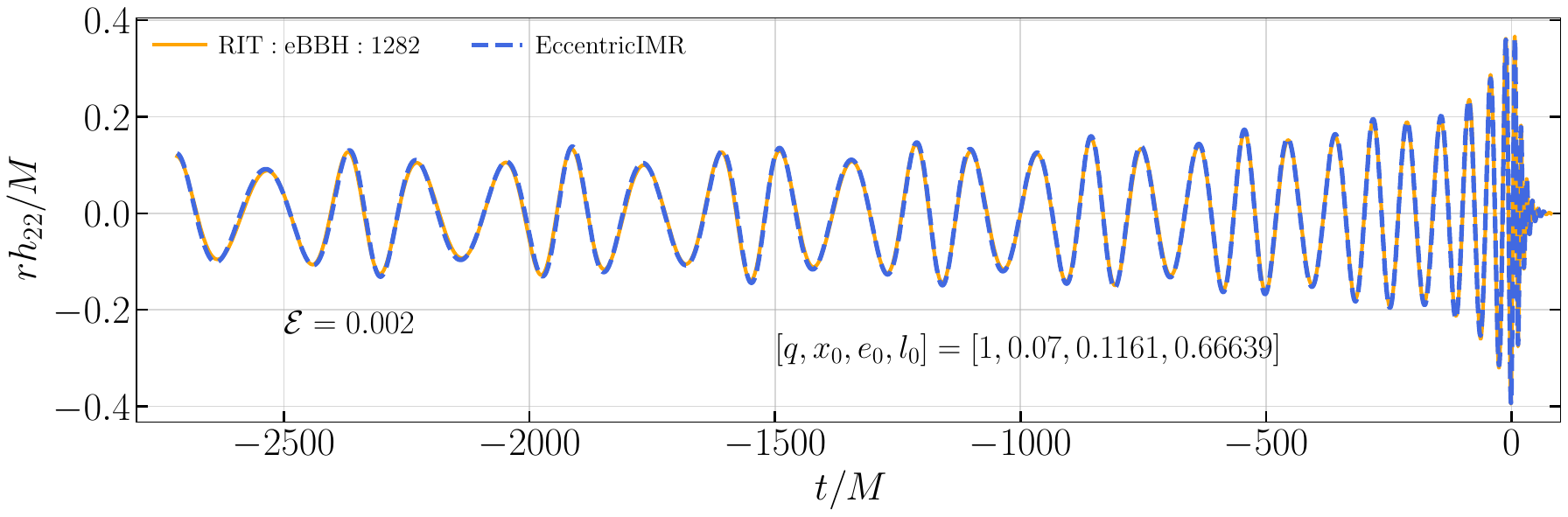}
		\label{fig:1282}
	}\\
	\subfloat[\texttt{RIT:eBBH:1422}]{
		\includegraphics[scale=0.5]{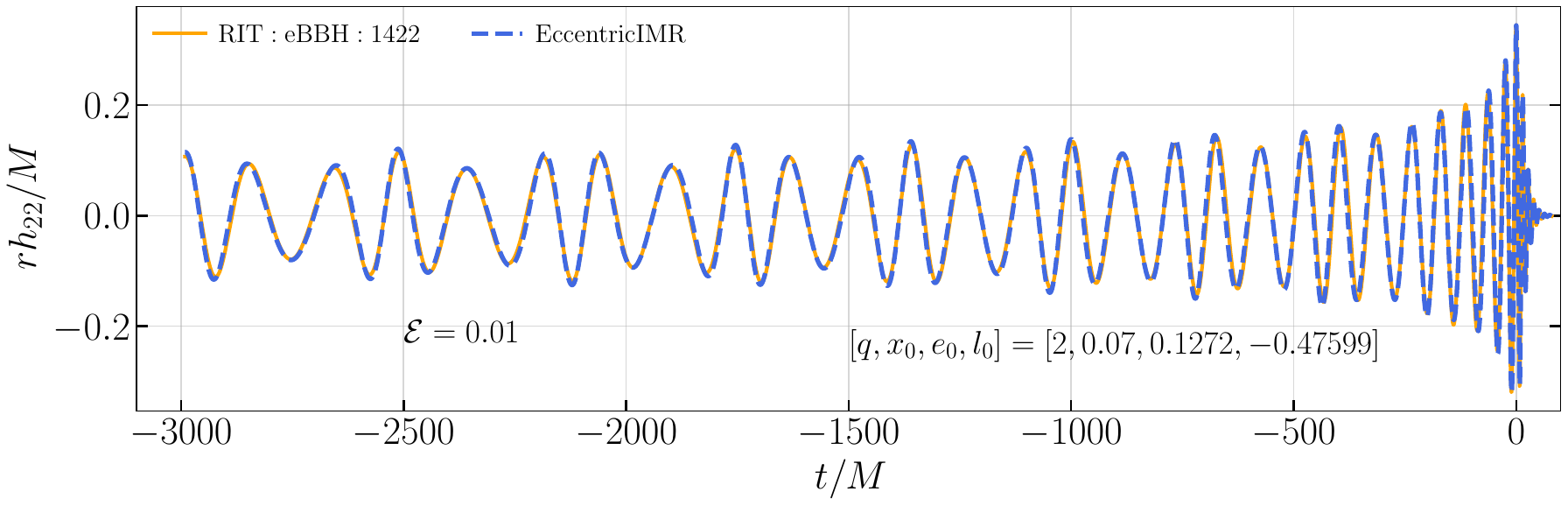}
		\label{fig:1422}
	}\\
    \subfloat[\texttt{RIT:eBBH:1468}]{
		\includegraphics[scale=0.5]{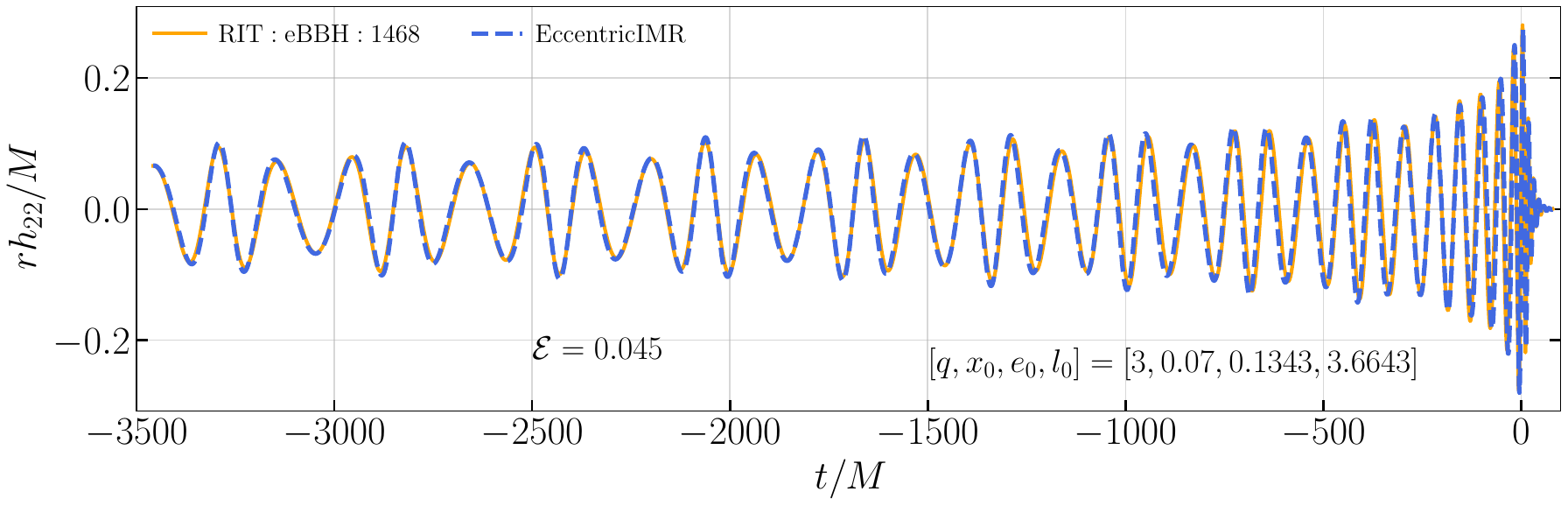}
		\label{fig:1468}
	}\\
	\subfloat[\texttt{RIT:eBBH:1491}]{
		\includegraphics[scale=0.5]{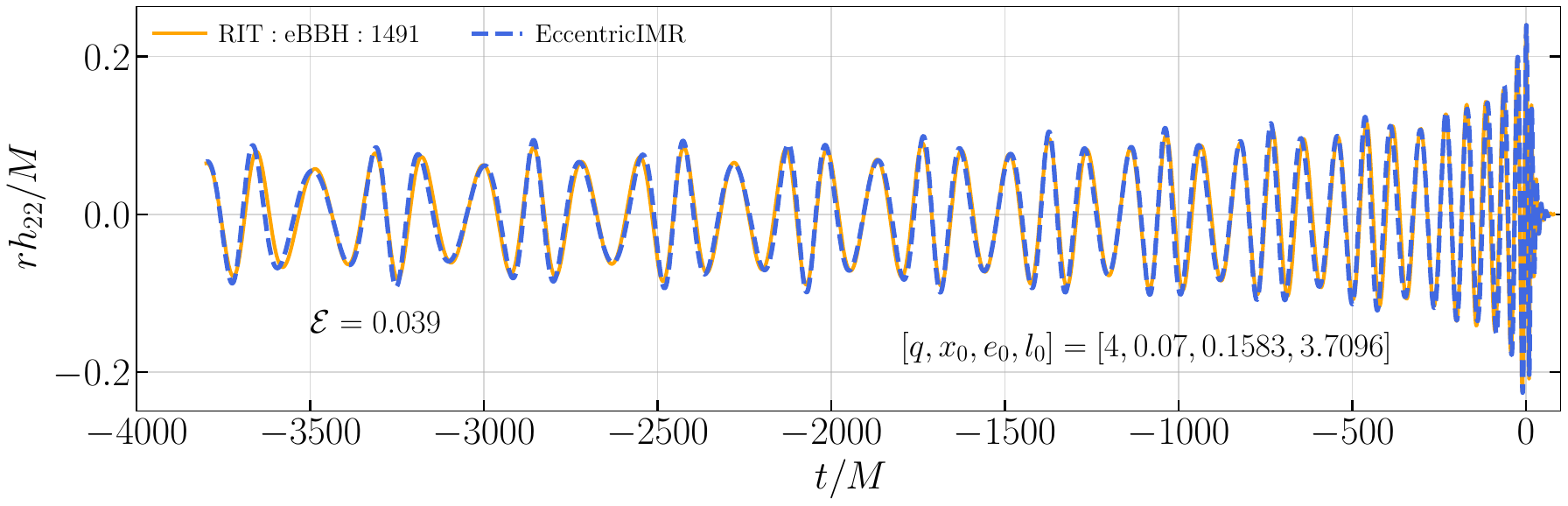}
		\label{fig:1491}
	}
\label{fig:RITNR}
\caption{We show four representative highly eccentric SXS NR simulations (blue dashed lines), namely, (a) \texttt{RIT:eBBH:1282}, (b) \texttt{RIT:eBBH:1422}, (c) \texttt{RIT:eBBH:1468} and (d) \texttt{RIT:eBBH:1491}, alongside the corresponding optimized \texttt{EccentricIMR} waveforms (orange solid lines). These simulations have eccentricities ranging from $e_0=0.12$ to $e_0=0.16$, given at a reference frequency of $x_0=0.07$. More details are in Section~\ref{sec:ritnr}.}
\end{figure*}

We then repeat the step for the remaining 14 SXS NR data and obtain reasonable agreement between NR and \texttt{EccentricIMR} model for most of the simulations. Figure~\ref{fig:SXSNR} shows three representative highly eccentric SXS NR simulations, namely, (a) \texttt{SXS:BBH:1361} ($q=1$), (b) \texttt{SXS:BBH:1368} ($q=2$), (c) \texttt{SXS:BBH:1372} ($q=3$) and (d) \texttt{SXS:BBH:1373} ($q=3$), alongside the corresponding optimized \texttt{EccentricIMR} waveforms. These simulations have eccentricities ranging from $e_0=0.132$ to $e_0=0.176$, given at a reference frequency of $x_0=0.07$. We find that the time-domain error between the NR data and optimized \texttt{EccentricIMR} waveforms are $0.01$, $0.016$, $0.04$ and $0.01$ respectively. In all cases, we see no visual differences between NR and optimized \texttt{EccentricIMR} waveforms. 

\subsection{Validation against RIT NR data}
\label{sec:ritnr}
Next, we proceed to the RIT data. Initially, we select a total of 23 non-spinning eccentric NR simulations with a minimum duration of approximately $\sim 1000M$. These simulations have quoted eccentricities up to $\sim 0.2$. We utilize all NR data up to $q=4$ for model validation, reserving the remaining 3 simulations to assess the validity of a circular merger model.

We observe that the \texttt{EccentricIMR} model provides a reasonable match to NR data for mass ratios up to $q\leq4$ and eccentricities up to $e=0.18$, defined at a reference frequency of $x=0.07$. Across most of the NR data, time-domain errors are approximately $\sim 10^{-2}$. However, in a few instances, errors escalate to around $\sim 10^{-1}$. To showcase the reasonable match of the \texttt{EccentricIMR} model against RIT NR data, we present four highly eccentric RIT NR simulations (\texttt{RIT:eBBH:1282}, \texttt{RIT:eBBH:1422}, \texttt{RIT:eBBH:1468}, and \texttt{RIT:eBBH:1491}) for mass ratios $q=[1,2,3,4]$, alongside their corresponding optimized \texttt{EccentricIMR} waveforms in Figure~\ref{fig:RITNR}. These simulations have eccentricities ranging from $e_0=0.12$ to $e_0=0.16$, given at a reference frequency of $x_0=0.07$. Notably, these represent the highest eccentric RIT simulations used in this study for each mass ratio value. It is important to also emphasize that there are no discernible visual differences between these two waveforms. The time-domain errors between NR data and \texttt{EccentricIMR} waveforms are $0.002$, $0.01$, $0.045$, and $0.039$ respectively.

\begin{figure}
\includegraphics[width=\columnwidth]{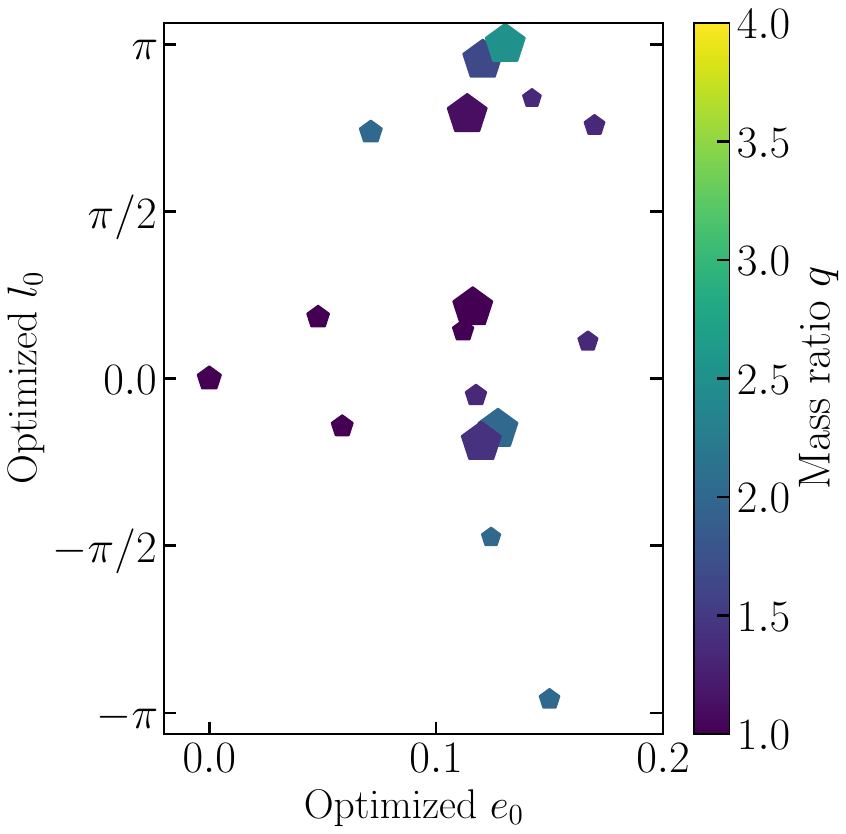}
\caption{We show the eccentricity $e_0$ and mean anomaly $l_0$ at a reference frequency $x_0=0.07$ for all NR-optimized \texttt{EccentricIMR} waveforms. Additionally, we color-code the eccentricity/mean anomaly values based on the mass ratios, and the marker size varies depending on the length of the NR data used. More details are in Section~\ref{sec:ritnr}.}
\label{fig:RITNRparams}
\end{figure}

In Figure~\ref{fig:RITNRparams}, we present the NR-optimized eccentricity $e_0$ and mean anomaly $l_0$ at a reference frequency $x_0=0.07$ for all 20 \texttt{EccentricIMR} waveforms. Additionally, we color-code the eccentricity/mean anomaly values based on the mass ratios. The marker size varies depending on the length of the NR data used. The longest NR data employed in our study is approximately $\sim 4000M$, while the shortest is about $\sim 1000M$ in duration. It is worth noting that most of the NR data falls between $q=1$ and $q=2$ (17 in total). Some of the high eccentric NR data also spans approximately $\sim 12000M$, but for simplicity, we utilize only the final $\sim 4000M$ of the binary evolution. Additionally, we observe that most of the NR data used in the validation yields optimizied eccentricities around $\sim 0.1$ and beyond.

\begin{figure*}
\includegraphics[width=\textwidth]{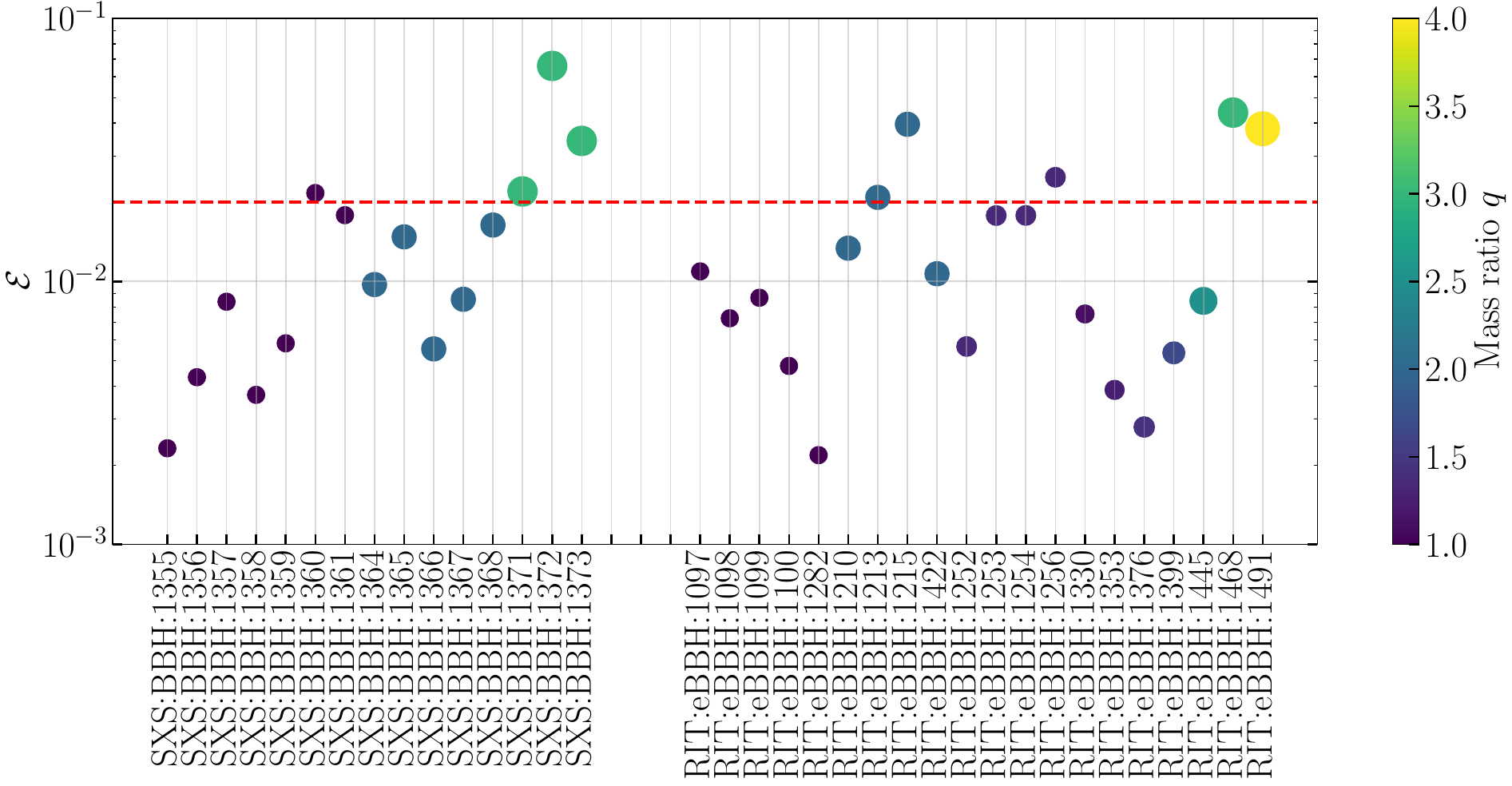}
\caption{We show the time-domain errors between SXS (RIT) NR data and optimized \texttt{EccentricIMR} waveforms. The error values are color-coded, and the markersize is assigned according to the mass ratio. More details are in Section~\ref{sec:sxs_rit_comparison}.}
\label{fig:mismatch}
\end{figure*}

\subsection{Comparison between SXS NR validation and RIT NR validation}
\label{sec:sxs_rit_comparison}
Now that we have demonstrated the \texttt{EccentricIMR} model's ability to reasonably match both SXS and RIT eccentric NR data, we proceed to quantify their accuracies and compare them against each other. 
Apart from NR validation of the \texttt{EccentricIMR} model, this also enables a cross-comparison between SXS and RIT NR data using the \texttt{EccentricIMR} model. 

\subsubsection{Comparison of the time-domain errors}
For each NR data used in validation, we calculate the time-domain time/phase optimized error.
We present the time-domain errors between NR data and optimized \texttt{EccentricIMR} waveforms in Figure~\ref{fig:mismatch}. Additionally, we colorcode the error values according to the mass ratio.
We observe that for most cases, time-domain errors are below $0.03$, often considered as a crude threshold for detection and source characterization readiness. 
We note a slight increase in errors as the binary moves away from the equal mass limit. 
Additionally, the time-domain errors obtained against the RIT NR data are more or less comparable to the ones obtained against the SXS NR data.

\begin{figure}
\includegraphics[width=\columnwidth]{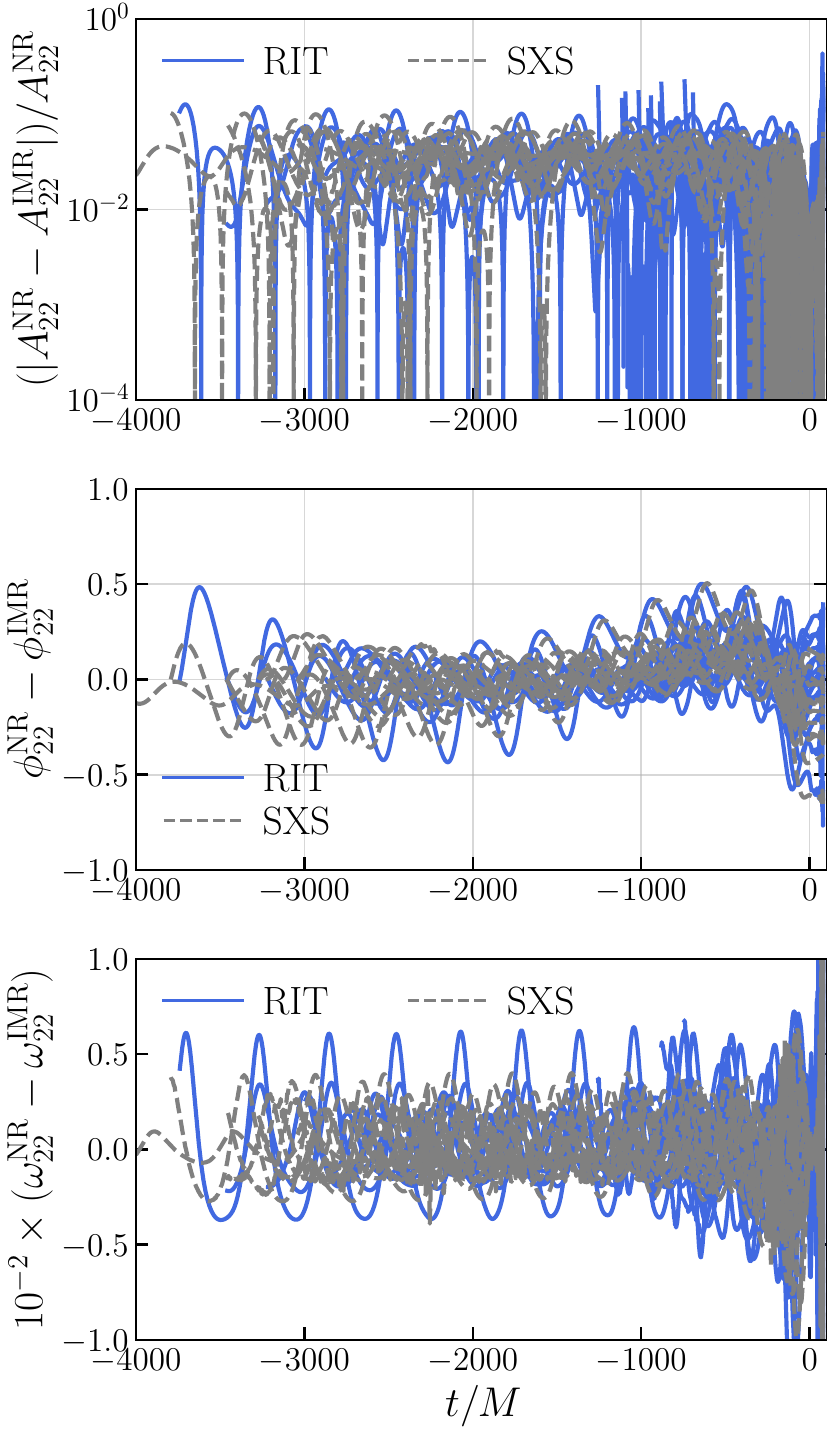}
\caption{We show relative amplitude errors (upper panel), absolute phase errors (middle panel) and the relative frequency errors (lower panel) between NR data and optimized \texttt{EccentricIMR} waveforms. Blue solid lines denote errors against RIT NR data while grey dashed lines are used for SXS NR data. More details are in Section~\ref{sec:sxs_rit_comparison}.}
\label{fig:NR_errors}
\end{figure}

\subsubsection{Comparison of the amplitudes, phases and frequencies}
Next, we calculate the relative amplitude difference
\begin{equation}
\frac{|A_{22}^{\tt NR} - A_{22}^{\tt IMR}|}{A_{22}^{\tt NR}},
\end{equation}
absolute phase difference
\begin{equation}
\phi_{22}^{\tt NR} - \phi_{22}^{\tt IMR},
\end{equation}
and relative $(2,2)$ mode frequency difference
\begin{equation}
\frac{\omega_{22}^{\tt NR} - \omega_{22}^{\tt IMR}}{\omega_{22}^{\tt NR}},
\end{equation}
between NR data and the corresponding optimized \texttt{EccentricIMR} waveform for all cases. Here, $A_{22}^{\rm NR}$ ($A_{22}^{\tt IMR}$), $\phi_{22}^{\rm NR}$ ($\phi_{22}^{\tt IMR}$), and $\omega_{22}^{\rm NR}$ ($\omega_{22}^{\tt IMR}$) represent the amplitude, phase, and instantaneous frequency of the NR data (\texttt{EccentricIMR} waveform), respectively. In Figure~\ref{fig:NR_errors}, we show these differences obtained using both SXS and RIT NR data and their corresponding \texttt{EccentricIMR} counterparts.

\begin{figure*}[htp]
	\subfloat[\texttt{MAYA:BBH:0927}]{
		\includegraphics[scale=0.5]{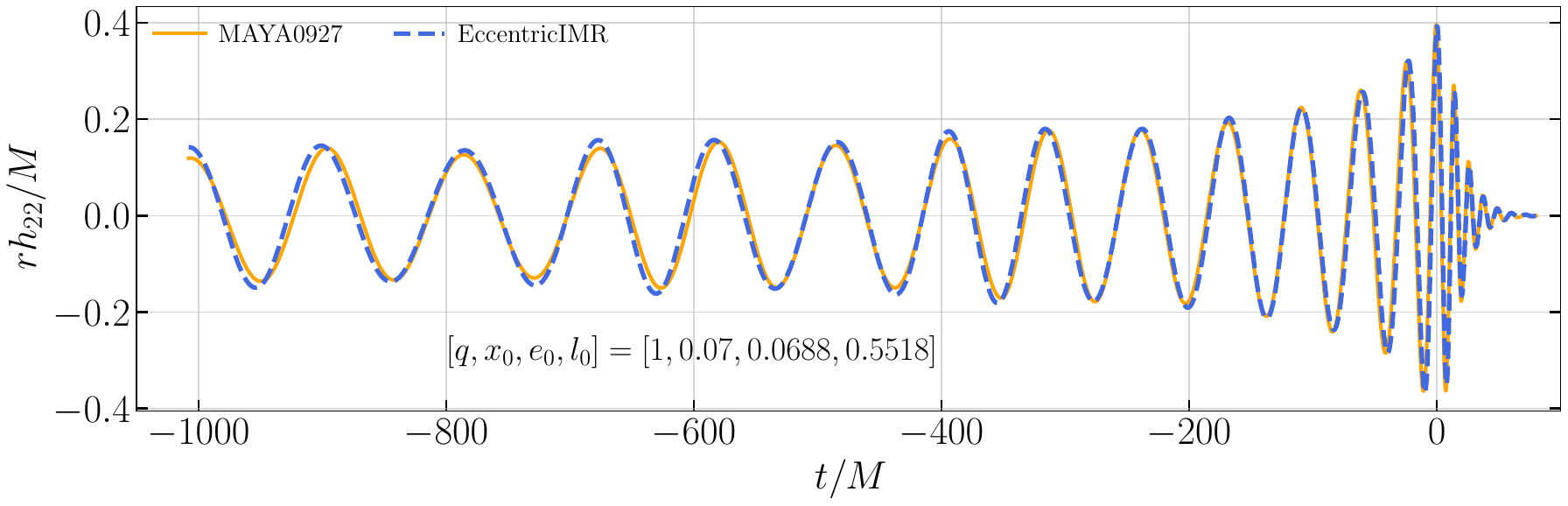}
		\label{fig:maya927}
	}\\
	\subfloat[\texttt{MAYA:BBH:0951}]{
		\includegraphics[scale=0.5]{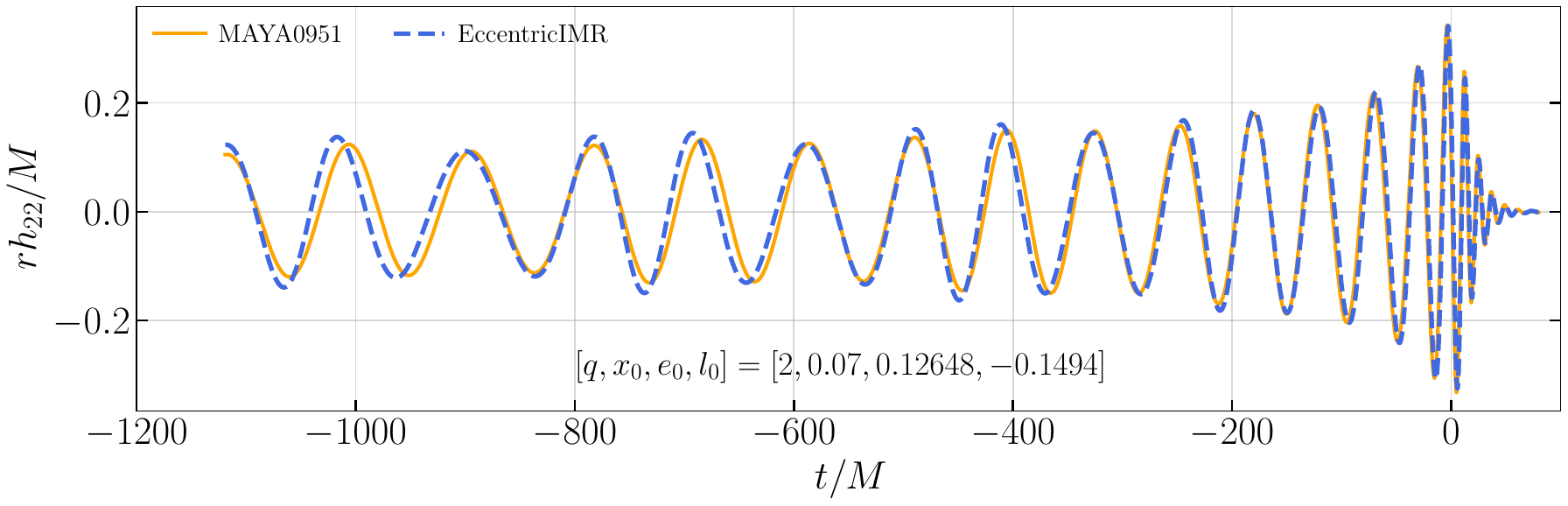}
		\label{fig:maya951}
	}
\label{fig:MAYANR}
\caption{We show two representative eccentric MAYA NR simulations (blue dashed lines), namely, (a) \texttt{MAYA:BBH:0927}, and (b) \texttt{MAYA:BBH:0951}, alongside the corresponding optimized \texttt{EccentricIMR} waveforms (orange solid lines). More details are in Section~\ref{sec:mayanr}.}
\end{figure*}

We observe that the relative amplitude differences are mostly $\sim0.02$, indicating that the amplitude errors for the \texttt{EccentricIMR} model are around $2$\%. Absolute phase errors are always subradian, regardless of whether we use SXS or RIT NR data for validation. Finally, frequency errors are always less than $1$\%. This implies that the \texttt{EccentricIMR} model can be effectively utilized to understand the phenomenology of eccentric BBH systems and in source characterization.

Another noteworthy observation is that the errors between SXS NR data and the \texttt{EccentricIMR} model are comparable to the errors between RIT NR data and the \texttt{EccentricIMR} model. Upon closer inspection, we find that errors between RIT NR data and the \texttt{EccentricIMR} model are slightly larger than the errors between SXS NR data and the \texttt{EccentricIMR} model (Figure~\ref{fig:NR_errors}). This corroborates our earlier finding that the overall time-domain errors between the \texttt{EccentricIMR} model and NR data are comparable between SXS and RIT data. This indicates that the eccentric NR simulations from the SXS collaboration and RIT have comparable accuracy for the $(2,2)$ mode.

Finally, we note that the differences between NR data and corresponding optimized \texttt{EccentricIMR} amplitudes, phases and frequencies exhibit systematic oscialltory behaviours. This is possibly due to higher order PN effects not included in the \texttt{EccentricIMR} model.

\subsection{Validation against MAYA NR data}
\label{sec:mayanr}

Apart from using SXS and RIT NR data, we also explore the potential of utilizing some of the recently available non-spinning eccentric NR simulations from the \texttt{MAYA} collaboration. These eccentric waveforms cover mass ratios up to $q=4$ and eccentricities up to $0.6$. However, these waveforms are considerably shorter in length, covering at most $\sim 1600M$ for the longest simulation. Most of the simulations are $\sim 1000M$ long or shorter. The substantially shorter duration of signals poses a challenge in utilizing them for model validation.

Nonetheless, we make an attempt to validate the \texttt{EccentricIMR} model against these NR simulations. Unlike the case for SXS or RIT data, we find that the model is unable to efficiently match these simulations. While some qualitative match has been achieved, the overall time-domain errors are relatively high - often larger than $0.5$. In Figure~\ref{fig:MAYANR}, we show two representative eccentric NR waveforms from the \texttt{MAYA} catalog with mass ratios $q=1$ and $q=2$ respectively along with corresponding optimized \texttt{EccentricIMR} waveforms. The quoted eccentricity value for both the NR simulations is $0.039$. We find that the optimized \texttt{EccentricIMR} waveform for these two cases have eccentricities $e_0=0.068$ and $e_0=0.126$ respectively at a reference frequency of $x_0=0.07$. We observe noticeable differences between NR and corresponding \texttt{EccentricIMR} waveforms. Similar differences between \texttt{EccentricIMR} and \texttt{MAYA} NR data are observed for other simulations too. Therefore, we do not include a detailed comparison here and only like to stress that while good quantitative agreement have not been achieved, we find reasonable qualitative matches between them.

\section{Validity of a circular merger model}
\label{sec:cmm}
One common modeling strategy for eccentric BBH waveforms is to employ an eccentric inspiral and attach a circular merger model to it - similar to the way \texttt{EccentricIMR} model is built. This treatment assumes that most of the eccentric binaries circularize sufficiently by the time they progress close to merger. However, this assumption has primarily been tested for binaries with mass ratios $q\leq3$ using SXS NR data. Here, we test this assumption for both the dominant $(2,2)$ mode and the subdominant modes for $q\leq7$ using SXS and RIT NR data

\subsection{Testing the circular merger model: $(2,2)$ mode}
\label{sec:cmm_22}
We focus on the dominant $(2,2)$ mode. We test the validity of a circular merger model in two different ways. First, we utilize recent eccentric RIT NR simulations, we re-evaluate the validity of this assumption for systems up to $q\leq7$. Second, we use the \texttt{EccentricIMR} model itself to understand the effectiveness of a circular merger model.

\subsubsection{Testing the circular merger model using RIT NR data}
\begin{figure*}[htp]
	\subfloat[]{\label{fig:cmm_q1to7}
		\includegraphics[scale=0.42]{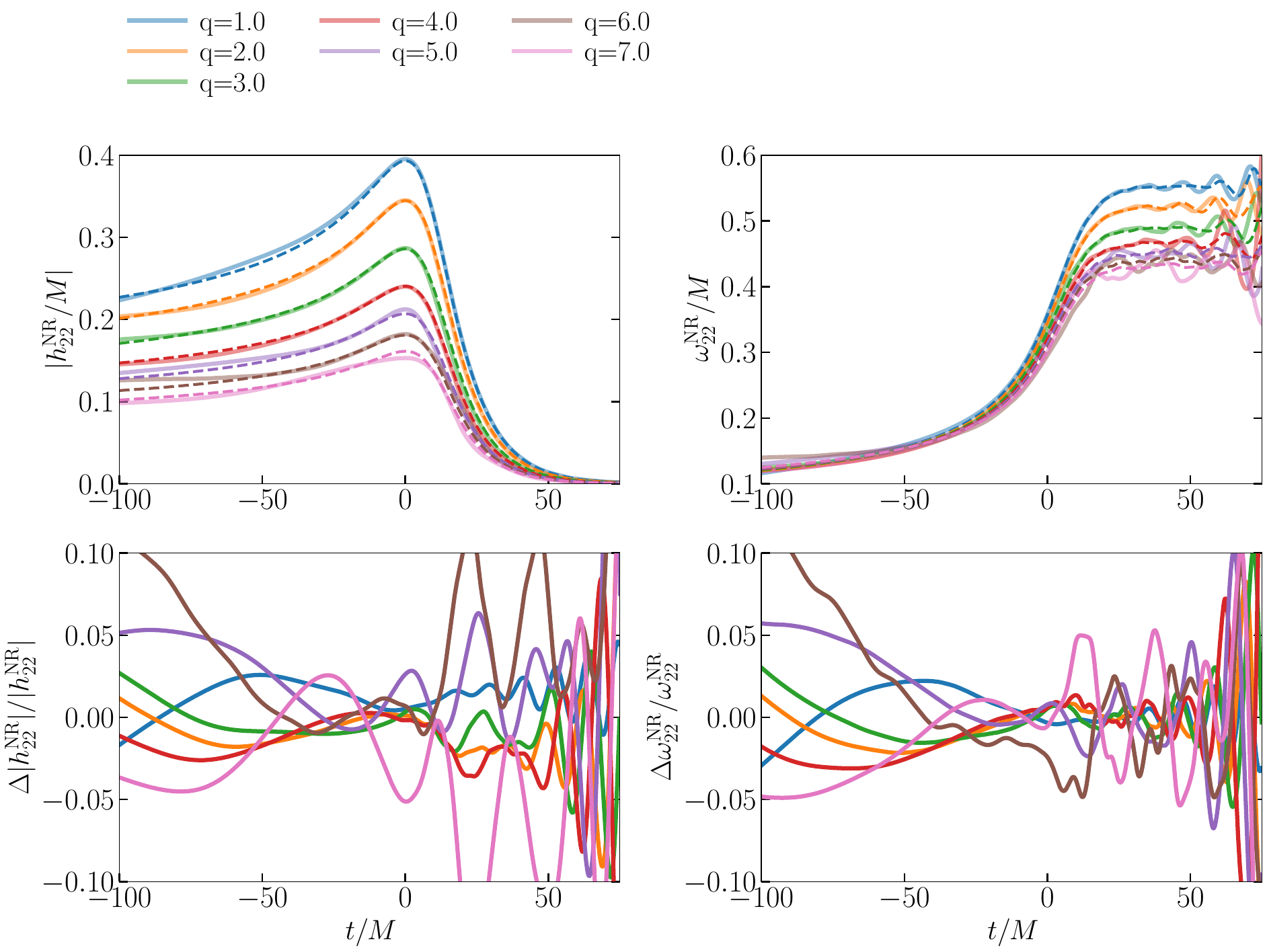}
	}\\
	\subfloat[]{\label{fig:cmm_q4}
		\includegraphics[scale=0.42]{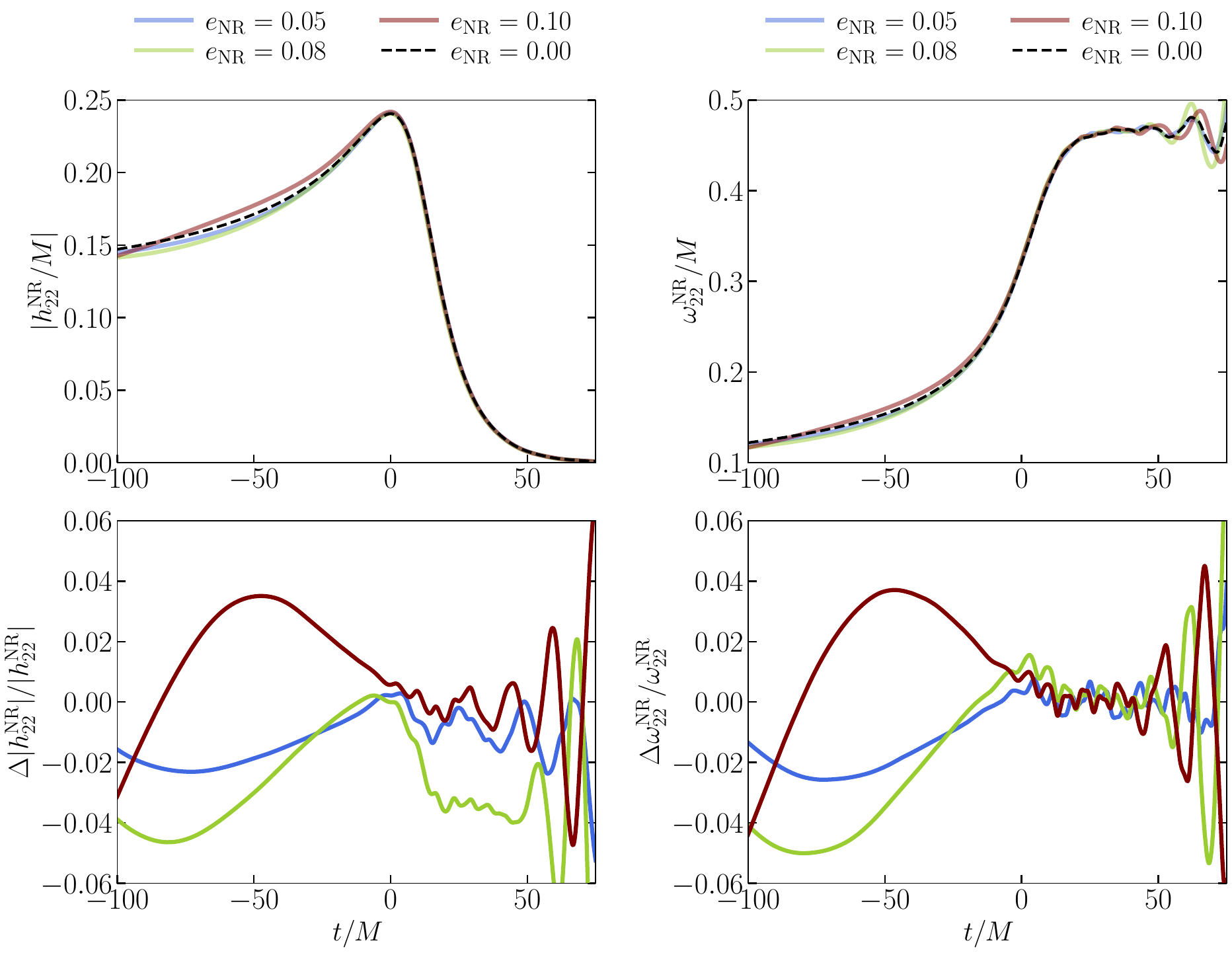}
	}\\
\caption{\label{fig:cmm}
(a) We show amplitudes (upper left panel) and frequencies (upper right panel) for highly eccentric NR waveforms for mass ratios $q=[1,2,3,4,5,6,7]$. 
(b) We show amplitudes (upper left panel) and frequencies (upper right panel) for NR waveforms for mass ratio $q=4$ with varying eccentricities up to quoted eccentricity of $0.19$. Additionally, we show the differences between eccentric and non-eccentric amplitudes and frequencies in lower panels for each case. More details are in Section~\ref{sec:cmm_22}.
}
\end{figure*}

We select seven highly eccentric NR simulations with quoted eccentricity of $0.19$ at the start of the waveform for mass ratios $q=[1,2,3,4,5,6,7]$. These simulations are \texttt{RIT:eBBH:1282}, \texttt{RIT:eBBH:1422}, \texttt{RIT:eBBH:1468}, \texttt{RIT:eBBH:1491}, \texttt{RIT:eBBH:1514}, \texttt{RIT:eBBH:1537} and \texttt{RIT:eBBH:1560} respectively.
In Figure~\ref{fig:cmm_q1to7}, we show the amplitudes and frequencies of these waveforms along with their non-eccentric counterparts. These non-eccentric counterparts are obtained from the following simulations respectively: 		\texttt{RIT:eBBH:1090}, \texttt{RIT:eBBH:1200}, \texttt{RIT:eBBH:0102}, \texttt{RIT:eBBH:1133}, \texttt{RIT:eBBH:0089}, \texttt{RIT:eBBH:0090} and \texttt{RIT:eBBH:0416}.
We find that for $t\ge-50M$ eccentric amplitude/frequencies align closely with the non-eccentric counterparts. To investigate it in more details, we also show the differences in amplitudes and frequencies between eccentric and non-eccentric waveforms. We find that these differences are around $1$\% to $10$\%. We note that the differences typically increase with mass ratio.

Next, we fix the mass ratio to $q=4$ and select three eccentric NR data (\texttt{RIT:eBBH:1143}, \texttt{RIT:eBBH:1149}, \texttt{RIT:eBBH:1153}) with quoted eccentricities of $0.05$, $0.08$, and $0.1$ respectively. Additionally, we choose the non-eccentric counterpart \texttt{RIT:eBBH:1133}. Figure~\ref{fig:cmm_q4} shows the amplitudes and frequencies of these waveforms. Furthermore, it also illustrates the corresponding differences between eccentric and non-eccentric amplitudes and frequencies. We find that the differences are mostly less than $6$\%. These differences typically increase with eccentricity. It is also interesting to observe that the differences in amplitudes and frequencies show similar qualitative features. 

Another interesting aspect of this investigation is that there are up to $10$\% differences in ringdown amplitudes and frequencies between eccentric and non-eccentric waveforms for all mass ratios. This indicates that the remnant mass and spin values of eccentric and non-eccentric BBHs are slightly different, and therefore, they exhibit slightly different ringdown structures. A more detailed study is necessary to understand this phenomenon.

\subsubsection{Testing the circular merger model using \texttt{EccentricIMR} model}
\begin{figure}
\includegraphics[width=\columnwidth]{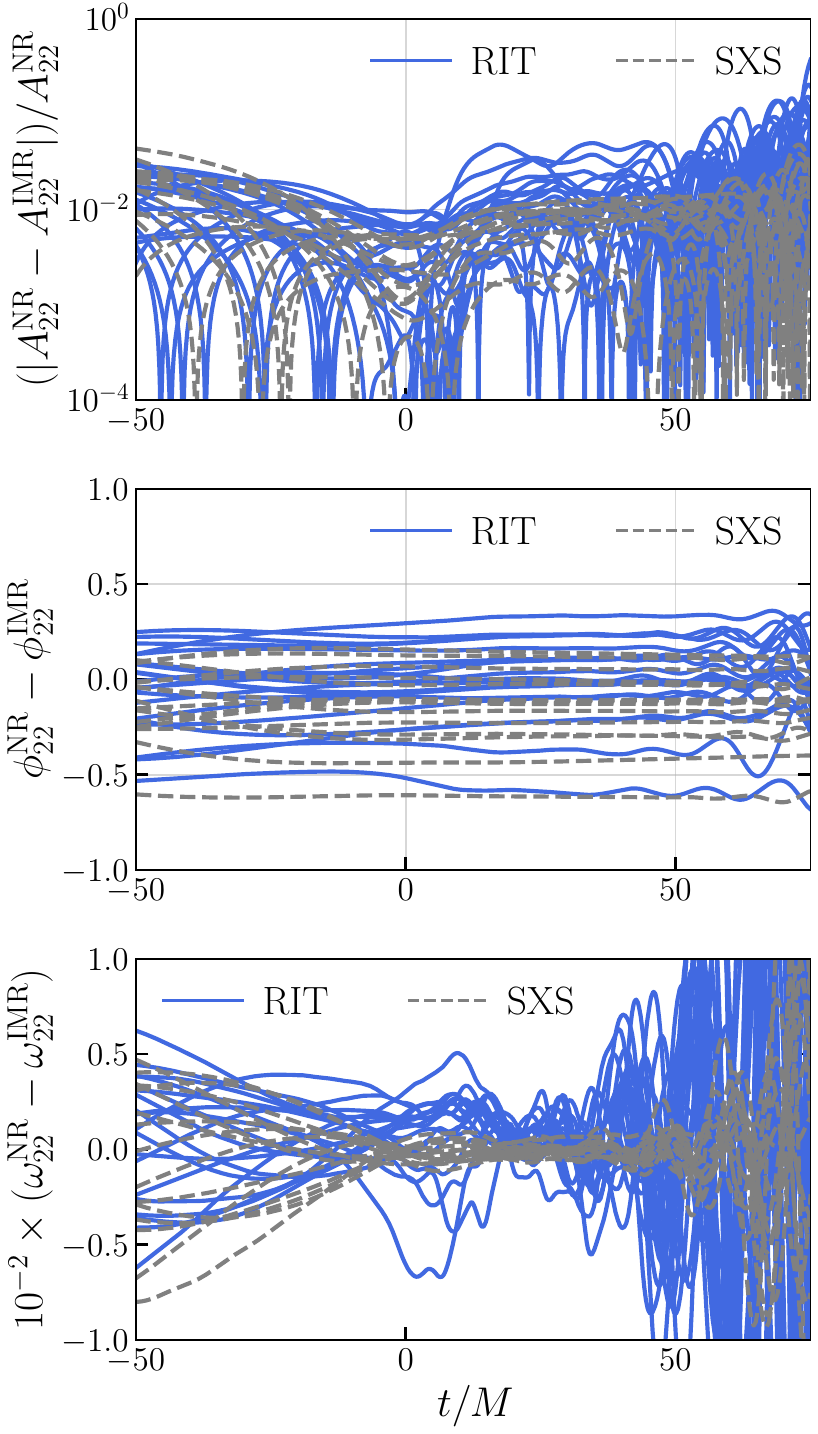}
\caption{We show relative amplitude errors (upper panel), absolute phase errors (middle panel) and the relative frequency errors (lower panel) between NR data and optimized \texttt{EccentricIMR} waveforms in the merger-ringdown part. Blue solid lines denote errors against RIT NR data while grey dashed lines are used for SXS NR data. More details are in Section~\ref{sec:cmm_22}.}
\label{fig:NR_errors_CMM}
\end{figure}

Another approach to assess the validity of the circular merger model involves examining the differences between NR data and the corresponding optimized \texttt{EccentricIMR} waveforms (obtained in Sections~\ref{sec:sxsnr} and \ref{sec:ritnr}) in the merger-ringdown part. In Fig.\ref{fig:NR_errors_CMM}, we present the differences in amplitudes, phases, and frequencies during the merger-ringdown phase. Our analysis indicates that the differences between \texttt{EccentricIMR} and NR phases consistently remain sub-radian. Regarding amplitudes, differences are approximately $1$\% up to $t=50M$, escalating to around $10^{-1}$ at $t=75M$. Frequency differences are confined within $0.2$\% and $0.5$\% for the SXS and RIT data, respectively, within the interval $-30M \leq t \leq 50M$. However, beyond $t=50M$, frequency errors undergo a rapid increase, reaching percent levels. This examination provides valuable insights into the behavior of the \texttt{EccentricIMR} model during the merger-ringdown phase, contributing to our understanding of its accuracy and the limitations of the circular merger approximation.

This analysis also indirectly suggests that while the numerical accuracy of the inspiral-merger-ringdown RIT NR waveforms are mostly comparable with the SXS NR waveforms for the $(2,2)$ mode, they may exhibit larger errors in the merger-ringdown part than the SXS NR data. It is noteworthy that, in Fig.\ref{fig:NR_errors}, only a couple of RIT NR simulations show larger differences than the SXS NR data when compared with optimized \texttt{EccentricIMR} waveforms. However, in Fig.\ref{fig:NR_errors_CMM}, almost for all cases, amplitude, phase, and frequency differences between RIT NR data and \texttt{EccentricIMR} waveforms are noticeably larger than in the case of SXS NR data. This observation highlights potential variations in the numerical accuracy of these NR simulations.

\begin{figure}
\includegraphics[width=\columnwidth]{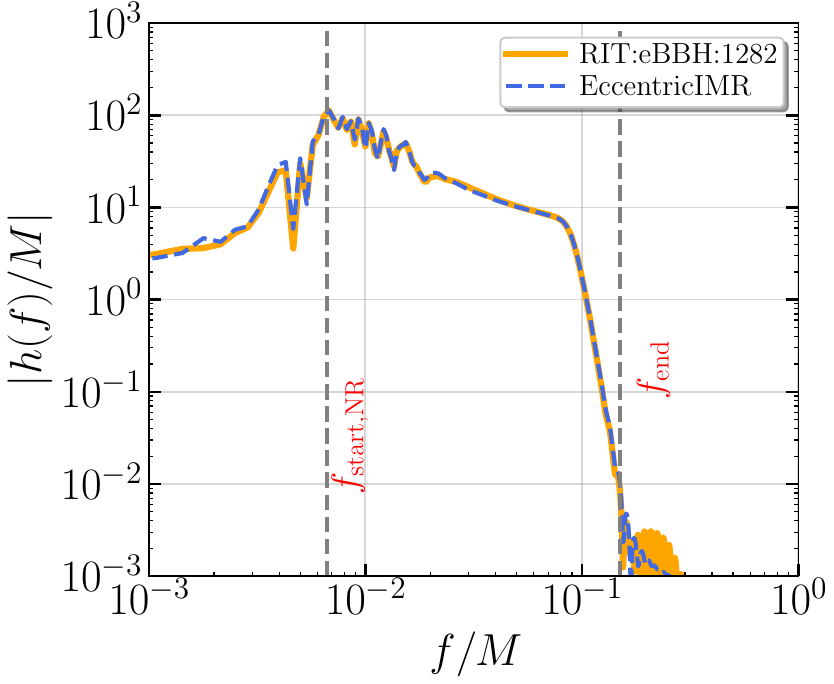}
\caption{We show the amplitude of the eccentric NR waveform \texttt{RIT:eBBH:1282} (orange solid line) characterized by $[q,\chi]:[1,0.19]$ and the optimized \texttt{EccentricIMR} waveform (blue dashed line) after transforming them into the frequency domain. Additionally, we indicate the initial starting frequency of the NR data computed directly from the time-domain waveform and the final frequency before it reaches the numerical noise floor as grey dashed vertical lines. More details are in Section~\ref{sec:cmm_22}.}
\label{fig:fd}
\end{figure}

\begin{figure*}
\includegraphics[scale=0.6]{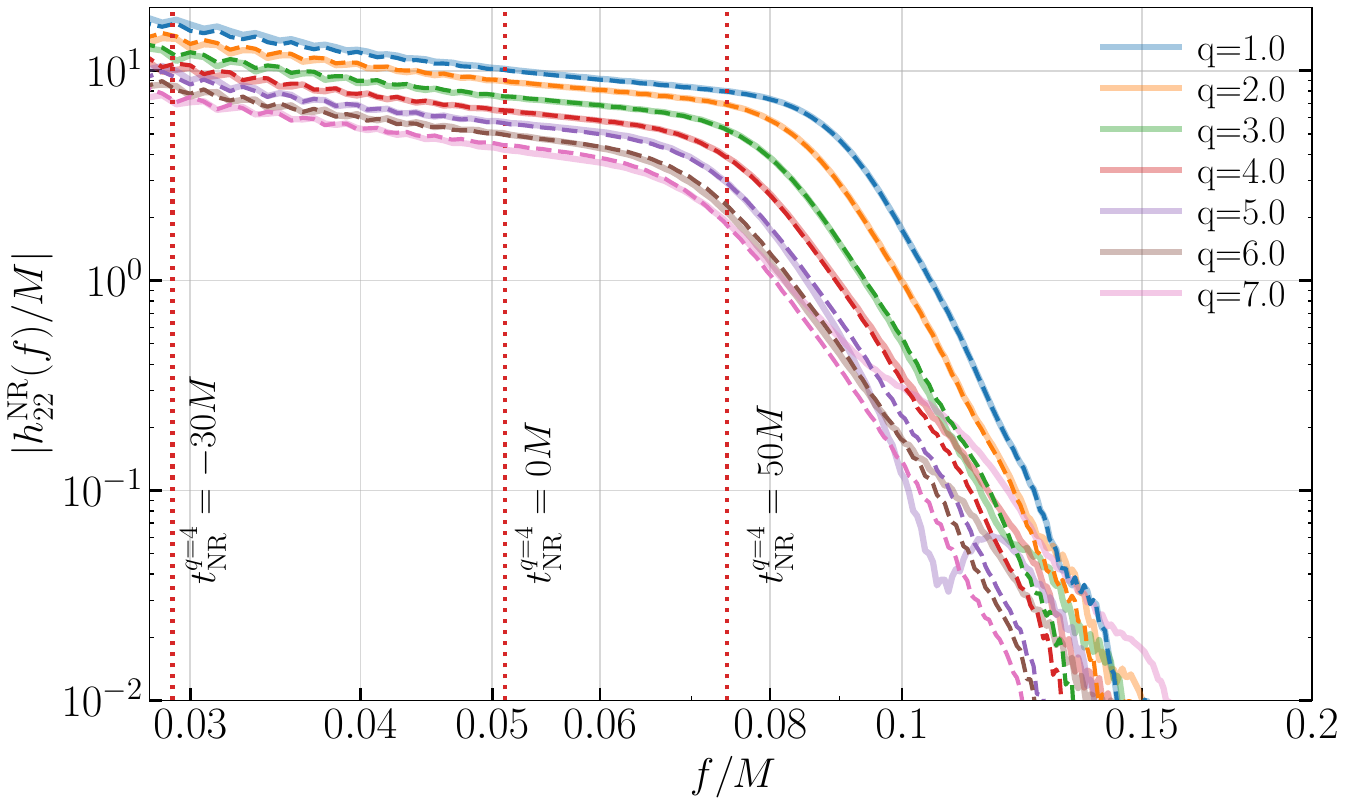}
\caption{We show frequency domain amplitudes for highly eccentric NR waveforms for mass ratios $q=[1,2,3,4,5,6,7]$ (solid lines). These NR simulations have quoted eccentricity of $\sim0.19$ at the start of the waveform. Additionally, we show the non-eccentric amplitudes as dashed lines. Red dashed vertical lines denote corresponding time for the non-eccentric NR simulations with $q=4$. More details are in Section~\ref{sec:cmm_22}.}
\label{fig:NR_FD_CMM}
\end{figure*}

\subsubsection{Testing the circular merger model in the frequency domain}
To gain further insights into the validity of the circular merger model, we analyze the waveforms in the frequency domain. 
However, it is crucial to note that additional complexities arise while obtaining a frequency domain eccentric waveform from its time-domain counterpart. 

To demonstrate these intricacies, in Figure~\ref{fig:fd}, we present the eccentric NR waveform \texttt{RIT:eBBH:1282}, characterized by $[q,e]=[1,0.19]$, and the optimized \texttt{EccentricIMR} waveform after transforming them into the frequency domain. We note that eccentricity introduces additional oscillatory modulations in the waveform. Separating out the modulations due to Gibbs phenomenon from those due to eccentricities in the initial portion of the waveform can be challenging. However, we adopt a simpler approach by identifying the frequency corresponding to the starting frequency of the time-domain waveform. We call this $f_{\rm start, NR}$. We discard frequencies smaller than this starting frequency. Additionally, we observe that after $f \geq 0.18M$, numerical errors in the NR data become dominant and call this $f_{\rm end}$. Therefore, we exclude any frequencies beyond $f_{\rm end}$. 

In Figure~\ref{fig:NR_FD_CMM}, we show the amplitude of eccentric (solid lines) and non-eccentric (dashed lines) waveforms for mass ratios ranging from $q=1$ to $q=7$. All the eccentric waveforms have a quoted NR eccentricity of $\sim0.19$ at the beginning. These simulations are \texttt{RIT:eBBH:1282}, \texttt{RIT:eBBH:1422}, \texttt{RIT:eBBH:1468}, \texttt{RIT:eBBH:1491}, \texttt{RIT:eBBH:1514}, \texttt{RIT:eBBH:1537} and \texttt{RIT:eBBH:1560} respectively.
Our focus is specifically on the merger-ringdown part. We therefore mark the frequencies corresponding to $t=-30M$, $t=0M$, and $t=50M$ for $q=4$ in a quasi-circular case. Although these values might vary slightly for other mass ratios, they provide a general overview. Our observations indicate that, in the merger-ringdown part, eccentric and non-eccentric amplitudes exhibit reasonable agreement in the regime $0.03 \leq f/M \leq 0.09$. However, at later times (i.e. higher frequencies) during the ringdown, eccentricity introduces additional features. These features may be associated with tail behaviors, and a more in-depth analysis is required for a comprehensive understanding.

\subsection{Testing the circular merger model: higher order modes}
\label{sec:cmm_hm}
\begin{figure*}
\includegraphics[width=\textwidth]{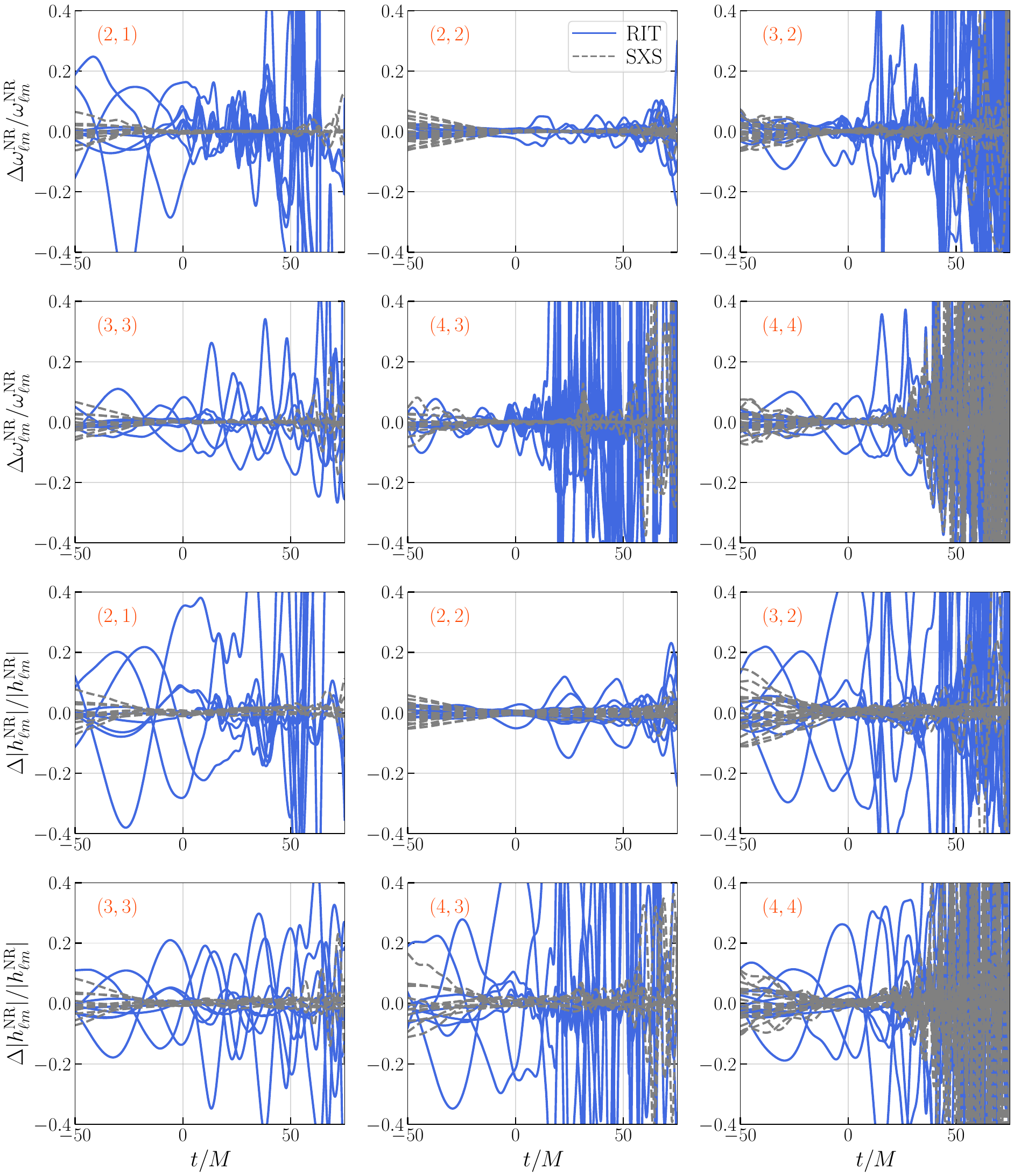}
\caption{We show relative amplitude errors (upper two panels) and relative frequency errors (lower two panels) between eccentric and corresponding non-eccentric NR waveforms in the merger-ringdown part for six representative modes: $[(2,1),(2,2),(3,2),(3,3),(4,3),(4,4)]$. Blue solid lines represent RIT NR data, while grey dashed lines depict SXS NR data. More details are in Section~\ref{sec:cmm_hm}.}
\label{fig:cmm_hm}
\end{figure*}

For completeness, we extend our analysis to higher-order modes, i.e., modes other than $(\ell,m)=(2,2)$. We utilize all of the NR simulations presented in Figure~\ref{fig:mismatch}. Figure~\ref{fig:cmm_hm} shows the relative differences in amplitudes and instantaneous frequencies between eccentric and corresponding non-eccentric NR waveforms in the merger-ringdown part forthe following modes: $[(2,1),(2,2),(3,2),(3,3),(4,3),(4,4)]$. 

The inclusion of the $(2,2)$ mode allows us to assess whether the differences become more pronounced as we move to higher-order modes. For the $q=1$ cases, we only calculate the differences for even $m$ modes as the odd $m$ modes are zero because of the symmetry. For SXS NR data, we observe that, for most modes, these differences are confined within $5\%$ in the time window $-30M \leq t \leq 50M$. Beyond $50M$, these differences escalate rapidly, likely due to increased numerical uncertainties at late times. Notably, these differences are more prominent in modes with $\ell \neq m$. 

Another noteworthy observation is the substantial increase in relative differences in amplitudes and instantaneous frequencies between eccentric and corresponding non-eccentric NR waveforms when using RIT NR data compared to SXS NR data. This discrepancy likely indicates that RIT NR data possesses larger numerical errors than the SXS data, particularly for higher modes. Notably, SXS NR data suggests that a circular merger model introduces only approximately $5\%$ errors in amplitudes and frequencies when compared against actual eccentric NR simulations. In contrast, RIT NR data would imply a much larger error. This underscores the significance of having high-accuracy NR data, particularly in the merger-ringdown regime.

\subsection{Testing the circular merger model: peak times of various modes}
\label{sec:cmm_peaks}
\begin{figure}
\includegraphics[width=\columnwidth]{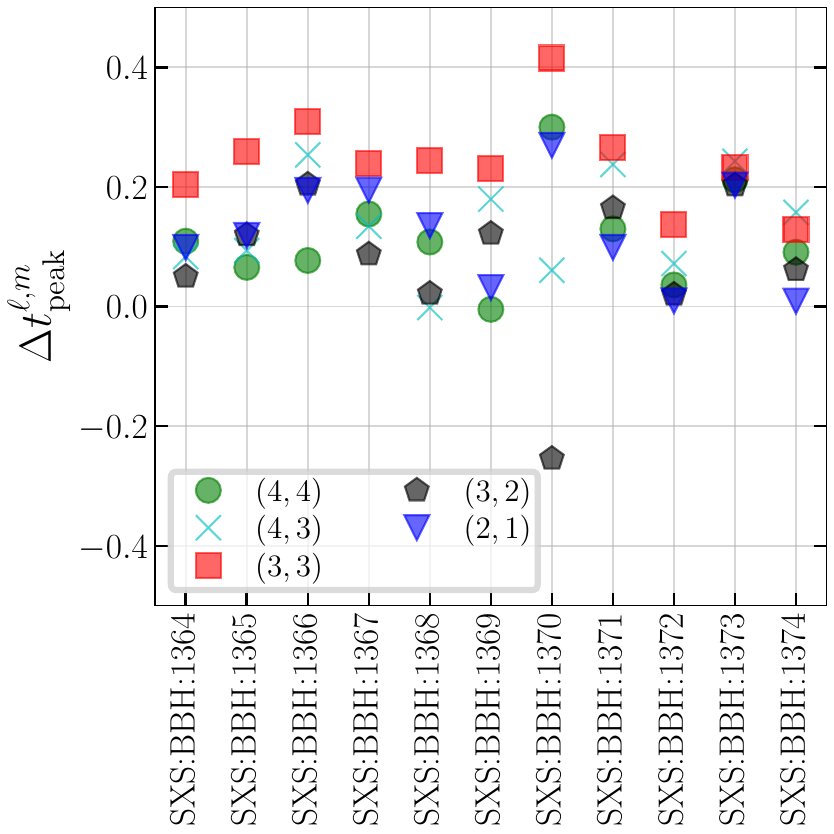}
\caption{We show peak time differences between eccentric and corresponding non-eccentric NR waveforms for six representative modes: $[(2,1),(3,2),(3,3),(4,3),(4,4)]$. More details are in Section~\ref{sec:cmm_peaks}.}
\label{fig:cmm_peak}
\end{figure}

We further investigate the peak times of different higher-order mode amplitudes to determine if eccentricity significantly affects them. We compute these peak times relative to the dominant $(2,2)$ mode's peak time. We restrict our analysis to SXS NR data for its smaller numerical errors compared to RIT NR data. We also focus only on $q=2$ and $q=3$ binaries as, for $q=1$ case, higher modes with odd values of $m$ will be zero due to the symmetry of the system. Figure~\ref{fig:cmm_peak} shows the peak time differences between eccentric and corresponding non-eccentric NR waveforms for six representative modes: $[(2,1),(3,2),(3,3),(4,3),(4,4)]$. The small peak time differences provide further support for the circular merger model in eccentric BBH waveform modeling.
 
\begin{figure}
\includegraphics[scale=0.58]{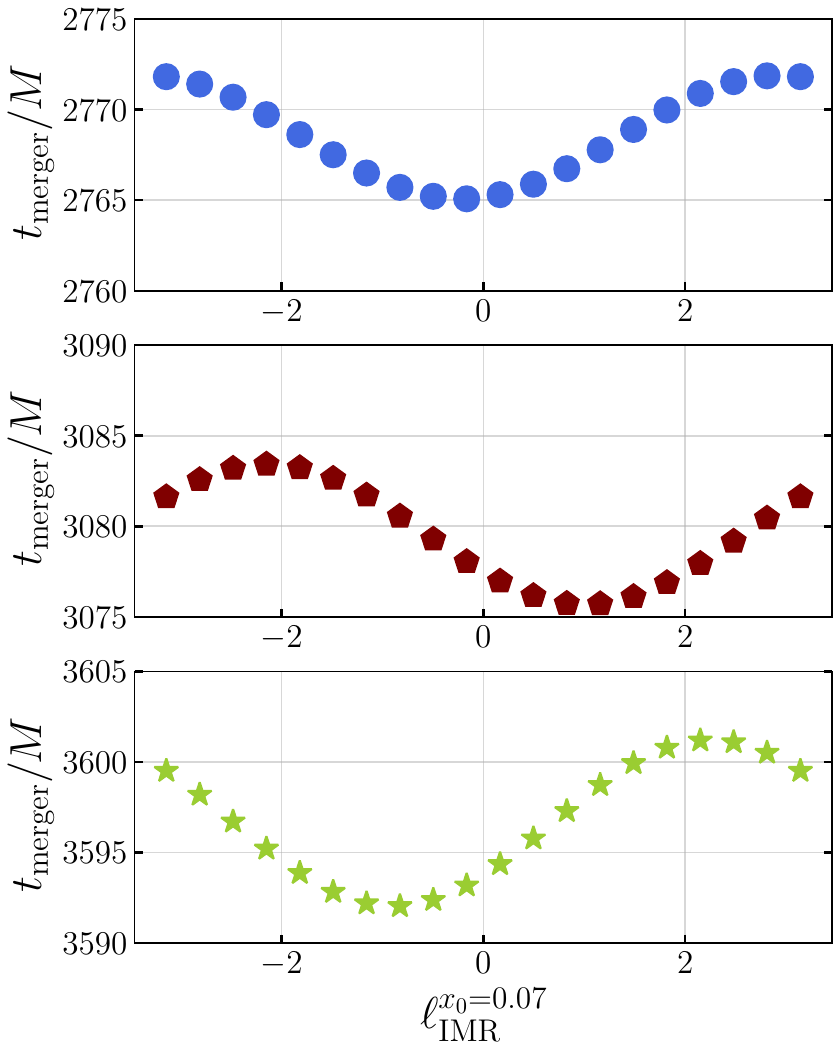}
\caption{We show the merger time as a function of the mean anomaly for $q=1$ (upper panel), $q=2$ (middle panel) and $q=3$ (lower panel). We set the eccentricity to be $e_{0}=1.0$ at a dimensionless frequency of $x_0=0.07$. More details are in Section~\ref{sec:merger_time}.}
\label{fig:tmerger_vs_meanano}
\end{figure}

\begin{figure}
\includegraphics[scale=0.58]{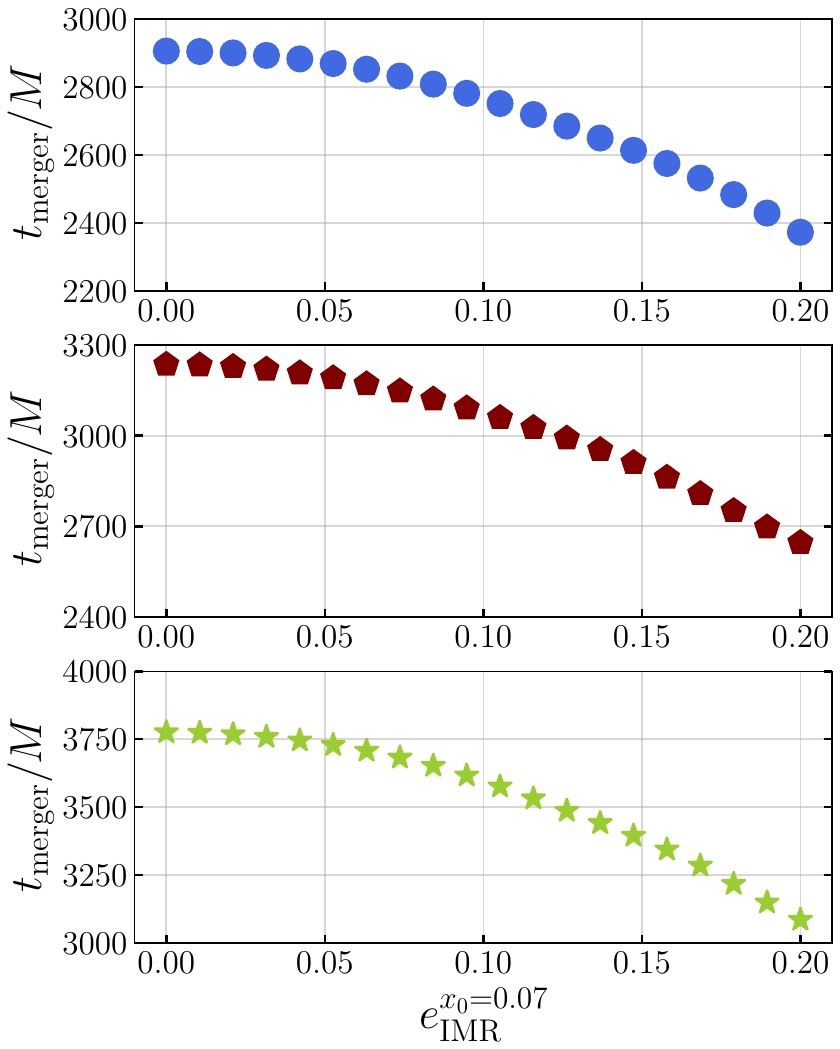}
\caption{We show the merger time as a function of the eccentricity for $q=1$ (upper panel), $q=2$ (middle panel) and $q=3$ (lower panel). We set the mean anomaly to be $l_{0}=0.0$ at a dimensionless frequency of $x_0=0.07$. More details are in Section~\ref{sec:merger_time}.}
\label{fig:tmerger_vs_ecc}
\end{figure}

\section{Phenomenology of eccentric BBH waveforms}
\label{sec:phenomenology}
Now that we have established a reasonable match between the \texttt{EccentricIMR} model and NR in Section~\ref{sec:nr_validation} and explored the validity of the circular merger model in Section~\ref{sec:cmm}, we proceed to utilize the \texttt{EccentricIMR} model to comprehend the phenomenology of eccentric binary black hole (BBH) waveforms.

\subsection{Understanding the merger time}
\label{sec:merger_time}
Our first investigation involves studying how the merger time in eccentric binaries varies with mean anomaly, eccentricity, and mass ratio. To conduct this analysis, we generate a set of waveforms at $q=1$, $q=2$, and $q=3$ for different eccentricities and mean anomalies at a reference dimensionless frequency of $x_0=0.07$. The merger time is computed as the time difference between the start of the waveform and the time when the amplitude of the $(2,2)$ mode peaks.

Figure~\ref{fig:tmerger_vs_meanano} shows the merger time as a function of the mean anomaly for different mass ratios while fixing the eccentricity at $e_{0}=0.1$. We find that the merger time shows a oscillatory dependence on the mean anomaly parameter. As the mass ratio increases, merger time increases too. Similarly, Figure~\ref{fig:tmerger_vs_ecc} shows the dependence of the merger time on the eccentricity. When the mean anomaly is held constant, we observe a monotonically decreasing trend in the merger time with increasing eccentricity.

\subsection{Understanding the effect of mean anomaly}
\label{sec:meanano}
\begin{figure*}[htp]
	\subfloat[]{
		\includegraphics[scale=0.45]{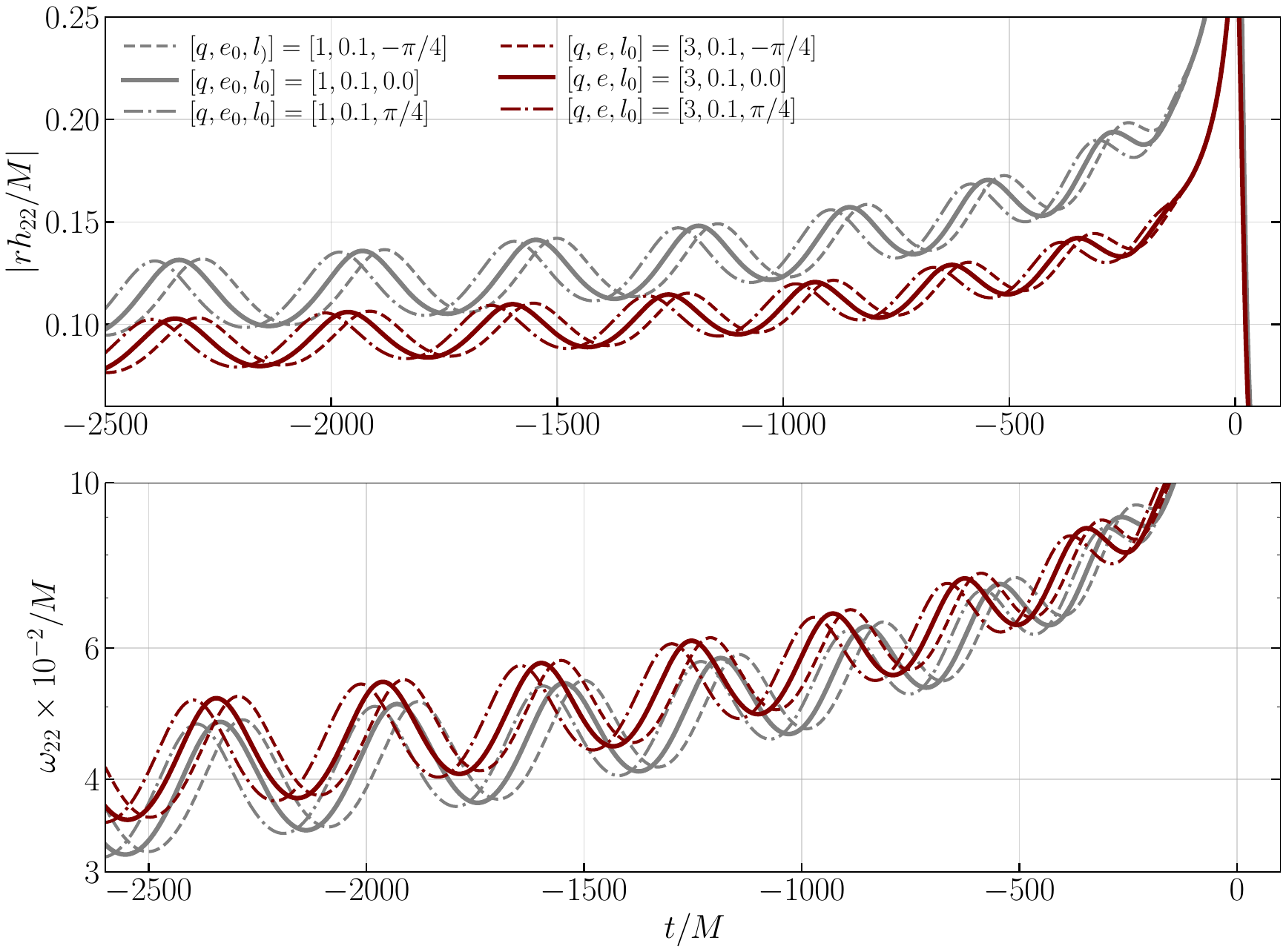}
		\label{fig:mean_ano_amp_freq}
	}\\
	\subfloat[]{
		\includegraphics[scale=0.45]{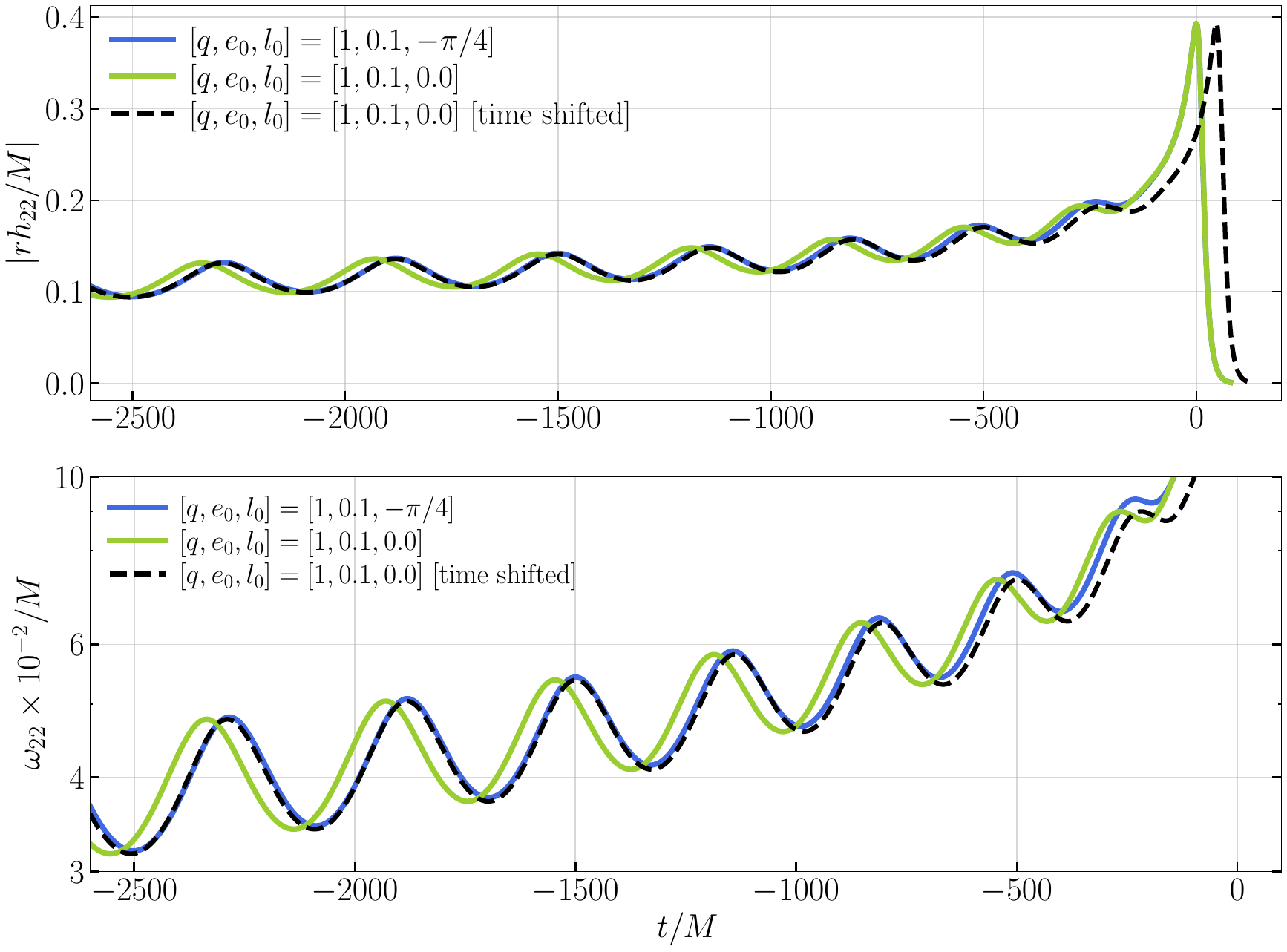}
		\label{fig:mean_ano_is_not_time_shift}
	}\\
\caption{\label{fig:mean_ano}
(a) We show the amplitudes (upper panel) and instantaneous frequencies (lower panel) of eccentric waveforms generated with \texttt{EccentricIMR} model for $q=1$ (grey lines) and $q=3$ (maroon lines) with three different mean anomaly values: $\ell=[-\pi/4,0,\pi/4]$.
(b) We show the amplitude and instantaneous frequencies of eccentric waveforms generated using the \texttt{EccentricIMR} model for $[q,e_0]=[1,0.1]$ with mean anomaly $l_0=0$ (represented by green solid lines) and $l=-\pi/4$ (depicted by blue solid lines). Additionally, we conduct a straightforward time-shift and present the resulting amplitude and frequencies as black dashed lines. More details are in Section~\ref{sec:meanano}.
}
\end{figure*}

Next, we investigate how mean anomaly affects the waveform morphology. We generate a series of eccentric waveforms utilizing the \texttt{EccentricIMR} model. These waveforms share the same mass ratio and eccentricity but differ in mean anomaly values. Figure \ref{fig:mean_ano_amp_freq} shows the amplitudes and instantaneous frequencies of eccentric waveforms for $q=1$ (depicted by grey lines) and $q=3$ (shown as maroon lines) across three distinct mean anomaly values: $l_0=[-\pi/4,0,\pi/4]$. 

As a naive guess, effect of mean anomaly may often be mistaken as a a simple time shift. In Figure~\ref{fig:mean_ano_is_not_time_shift}, we investigate whether a simple time shift can effectively compensate for the impact of mean anomaly. We generate eccentric waveforms with $[q,e_0]=[1,0.1]$ for mean anomaly $l_0=0$ and $l_0=-\pi/4$. In all cases, we define the eccentricities at a dimensionless reference frequency of $x_0=0.07$. The results indicate that a simple time shift is adequate to undo the effect of mean anomaly in the inspiral part. However, this shifts the merger time of the binary. A simple time shift is therefore not sufficient to mimic the effect of mean anomaly in inspiral-merger-ringdown waveforms.

\subsection{Understanding eccentric parameter space}
\label{sec:eccparamspace}
\begin{figure}[htp]
	\subfloat[]{
		\includegraphics[width=\columnwidth]{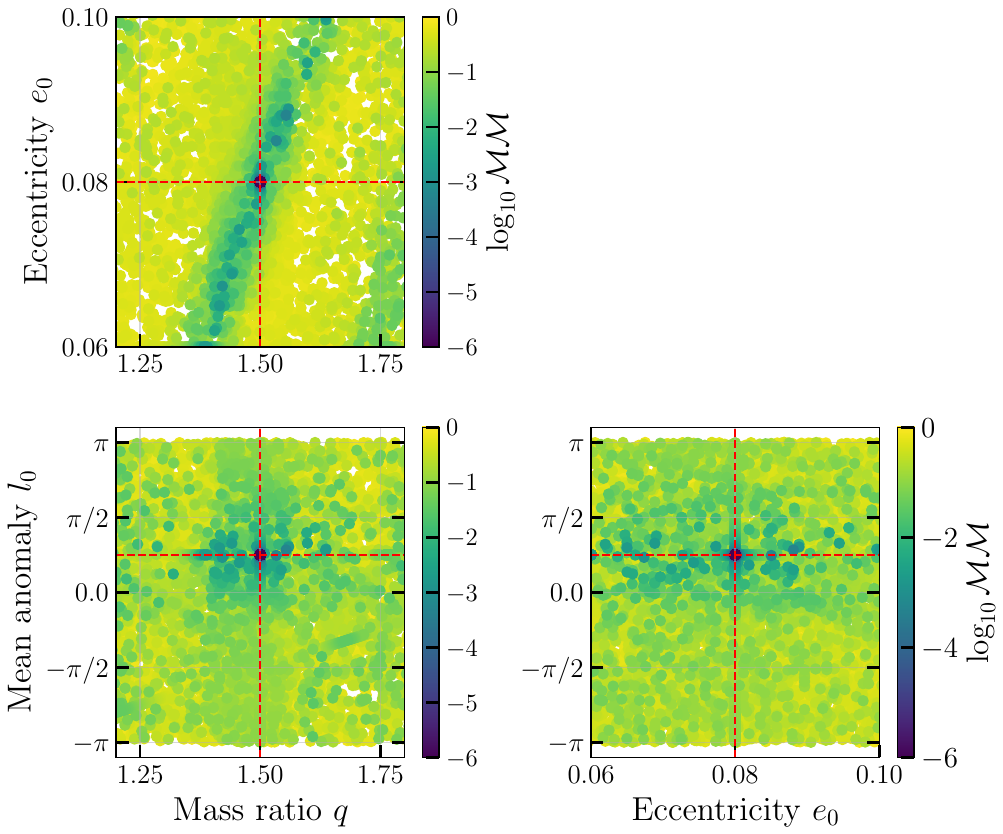}
        \label{fig:mismatch_param_space_zoomed}
        	}\\
	\subfloat[]{
		\includegraphics[scale=0.45]{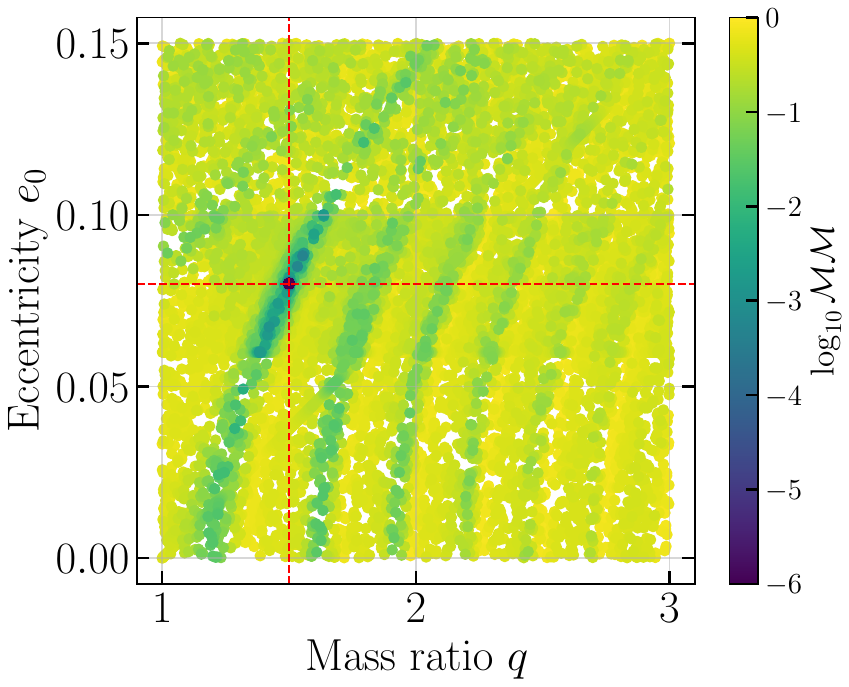}
		\label{fig:mismatch_ecc_q}
	}\\
\caption{
(a) We show the mismatches between reference waveform with $[q,e_0,l_0]=[1,0.08,\pi/4]$ and waveforms generated at different mass ratio, eccentricity, mean anomaly values around the parameters associated with the reference waveform. (b) Mismatches as a function of mass ratio and eccentricity for the entire parameter space. More details are in Section~\ref{sec:eccparamspace}.
}
\end{figure}

To explore how waveforms change in the eccentric parameter space, we conduct a mismatch study. We generate a reference waveform, denoted as $h_{\rm ref}$, with $[q,e_0,l_0]=[1,0.08,\pi/4]$. Subsequently, we compute mismatches between $h_{\rm ref}$ and waveforms generated with random mass ratios ($q$ in the range $[1,3]$), eccentricities ($e_0$ in the range $[0.0,0.15]$), and mean anomalies ($l_0$ in the range $[-\pi,\pi]$) (Figure~\ref{fig:mismatch_param_space_zoomed}). 

We calculate the frequency-domain mismatches assuming a flat noise curve. Frequency domain mismatch between two \texttt{EccentricIMR} waveforms $h_1$ and $h_2$ is defined as:
\begin{gather}
	\mathcal{M} \mathcal{M} = 1 - 4 \mathrm{Re}
	\int_{f_{\mathrm{min}}}^{f_{\mathrm{max}}}
	\frac{\tilde{h}_{1} (f) \tilde{h}_{2}^* (f) }{S_n (f)} df,
	\label{Eq:freq_domain_Mismatch}
\end{gather}
where $\tilde{h}(f)$ indicates the Fourier transform of the complex strain $h(t)$, $^*$ indicates complex conjugation, `$\mathrm{Re}$' indicates the real part, and $S_n(f)$ is the flat-noise curve. We set $f_{\rm min}$ and $f_{\rm max}$ to be the minimum and maximum frequency of the time domain data. While computing the mismatches, we only use the frequencies that fall within $f=[f_{\rm start, NR},f_{\rm end}]$. Before transforming the time domain waveforms to the frequency domain, we first taper the time domain waveform using a Planck window~\cite{McKechan:2010kp}, and then zero-pad to the nearest power of two. The mismatches are always optimized over shifts in time, polarization angle, and initial orbital phase.

We opt to compute only the frequency domain mismatches using a flat-noise curve, without incorporating a LIGO sensitivity curve. This choice is made to ensure a waveform-level comparison without the influence of their anticipated signal-to-noise ratios in any detector.
We find that, unsurprisingly, the minimum mismatch corresponds to the eccentricity, mean anomaly and mass ratio value of the reference waveform (Figure~\ref{fig:mismatch_param_space_zoomed}). Furthermore, mismatches increase rapidly as the eccentricity and mass ratio values deviate from the reference parameters. Figure~\ref{fig:mismatch_ecc_q} shows the mismatches as a function of mass ratio and eccentricity for the entire parameter space i.e. for $q=[1,3]$ and $e_0=[0,0.15]$. We notice that mismatches shows multiple bands of local minima indicating possible slight degeneracy  between eccentricity and mass ratio values.

\section{Discussion \& concluding remarks}
\label{sec:discussion}
In this paper, we revisit one of the existing time-domain eccentric IMR waveform models, namely \texttt{EccentricIMR}. The model is built by blending PN inspiral approximation with a circular merger model and can only generate the dominant $(2,2)$ mode waveform. The model uses a two-parameter characterization (via eccentricity and mean anomaly) of the eccentric waveforms, unlike most other available models that use a single eccentricity parameter. First, we introduce an efficient Python wrapper for the original \texttt{Mathematica} package, aiming to facilitate the utilization of this model in targeted search and data analysis studies. We find that the typical waveofrm generation cost is $\sim2s$.

Next, we identify a potential issue in the original NR validation of the model and propose an alternative strategy. Instead of matching the PN inspiral to NR data close to the merger and obtaining eccentricity/mean anomaly values there, we generate the waveform with eccentricity/mean anomaly values quoted much earlier in the binary inspiral. Subsequently, we match the entire available NR data.

Using the aforementioned strategy, we first re-validate the model against the 15 SXS NR data, used in original validation, covering mass ratios ranging from $q=1$ to $q=3$ and eccentricities up to $\sim 0.2$, estimated approximately 20 cycles before the merger. We find that our validation strategy yields similar accuracy for the \texttt{EccentricIMR} model. For most of the NR data, overall time-domain errors are less than $0.03$, indicating a reasonable match. In cases where the errors are slightly larger, we do notice significant qualitative agreement. This is assuring that the model is realistic and can be used in understanding phenomenology of eccentric waveforms.

Next, we validate the model against a set of 20 recently available RIT NR data, covering mass ratios up to $q=4$ and quoted eccentricities up to $0.19$. We find that overall time-domain errors obtained in this case are quite similar to the values obtained when SXS NR data are used. Furthermore, differences between NR data and corresponding \texttt{EccentricIMR} waveforms in amplitude, phase, and frequencies are overall comparable in both cases. Amplitude and frequency differences are mostly $\sim2$\% while phase errors are always sub-radian.
This implies that the \texttt{EccentricIMR} model is quite trustworthy as long as SXS NR and RIT NR data are concerned. Furthermore, it also means that the eccentric NR simulations from the SXS collaboration and the RIT group are quite similar to each other. This study therefore provides an alternative way to cross-compare RIT and SXS eccentric NR simulations for the $(2,2)$ mode. 

We further explore the possibility of validating the model against eccentric NR data obtained from the \texttt{MAYA} catalog. While we observe qualitative agreement between these NR data and \texttt{EccentricIMR} model, overall time-domain errors for these cases are larger than the ones obtained using SXS NR or RIT NR data.

The next piece of our investigation involves examining the validity of a circular merger approximation for eccentric binaries. Using RIT NR data up to $q=7$, we find that the differences between eccentric and non-eccentric waveforms in the merger-ringdown part can vary from $1$\% to $10$\%, depending on the mass ratio and eccentricity of the system. Differences rise when the value of either of these two quantities increases. We then study the validity of the circular merger model for the higher order modes using both RIT and SXS data. Our results suggest that RIT NR data has larger errors in the merger-ringdown part compared to SXS NR data. When SXS NR data are used, differences between eccentric and non-eccentric waveform amplitudes and instantaneous frequencies are always within $5\%$ indicating the effectiveness of circular merger models. We also notice that eccentricity does not significantly alter the relative peak times for different modes (compared to the dominant $(2,2)$ mode).

This work also helps us to reassess the accuracy of the \texttt{EccentricIMR} model and PN approximations in light of new eccentric NR simulations so that we can use \texttt{EccentricIMR} model to study the phenomenology of eccentric BBH waveforms (which we do in Section~\ref{sec:phenomenology}). In particular, we study how mean anomaly affect waveform morphology and merger time. We find that while, for inspiral-only waveforms, effect of mean anomaly can be mimicked by a time-shift, such an intuition is no longer valid for full inspiral-merger-ringdown waveform. Performing a mismatch study, we demonstrate that mean anomaly parameter is important to correctly recover the injected reference waveform. We further show that there exist a weak degeneracy between mass ratio and eccentricity.

Our study carries significant implications on the comparison between NR and PN approximations in eccentric BBH mergers. The reasonable accuracy observed between NR and the \texttt{EccentricIMR} model suggests that PN approximations can be reliably extrapolated close to the merger to construct accurate waveform models in eccentric BBH mergers. Potential drawback of the \texttt{EccentricIMR} model is the waveform generation time and unavailability of higher order modes. While this model is quite trustworthy when it comes to SXS and RIT eccentric NR data, it is not reliable for real-time data analysis unless significant computing resources are available. It would be beneficial to develop an efficient reduced-order surrogate representation of this model to reduce the waveform generation cost. We leave this work for future.

We believe that our findings will be helpful in developing phenomenological mulit-modal eccentric BBH waveform models in near future by combining PN approximations for the inspiral and a circular merger model for merger-ringdown. 

\begin{acknowledgments}
We thank Ian Hinder, Lawrence E. Kidder, and Harald P. Pfeiffer for making \texttt{EccentricIMR} model publicly available. We are also grateful to Scott Field, Vijay Varma, Gaurav Khanna, Ajit Mehta and Tejaswi Venumadhav for useful discussion. We thank the SXS collaboration, MAYA collaboration and RIT NR group for maintaining publicly available catalog of NR simulations which has been used in this study.
\end{acknowledgments}

\bibliography{References}

\begin{thebibliography}{39}%
\makeatletter
\providecommand \@ifxundefined [1]{%
 \@ifx{#1\undefined}
}%
\providecommand \@ifnum [1]{%
 \ifnum #1\expandafter \@firstoftwo
 \else \expandafter \@secondoftwo
 \fi
}%
\providecommand \@ifx [1]{%
 \ifx #1\expandafter \@firstoftwo
 \else \expandafter \@secondoftwo
 \fi
}%
\providecommand \natexlab [1]{#1}%
\providecommand \enquote  [1]{``#1''}%
\providecommand \bibnamefont  [1]{#1}%
\providecommand \bibfnamefont [1]{#1}%
\providecommand \citenamefont [1]{#1}%
\providecommand \href@noop [0]{\@secondoftwo}%
\providecommand \href [0]{\begingroup \@sanitize@url \@href}%
\providecommand \@href[1]{\@@startlink{#1}\@@href}%
\providecommand \@@href[1]{\endgroup#1\@@endlink}%
\providecommand \@sanitize@url [0]{\catcode `\\12\catcode `\$12\catcode `\&12\catcode `\#12\catcode `\^12\catcode `\_12\catcode `\%12\relax}%
\providecommand \@@startlink[1]{}%
\providecommand \@@endlink[0]{}%
\providecommand \url  [0]{\begingroup\@sanitize@url \@url }%
\providecommand \@url [1]{\endgroup\@href {#1}{\urlprefix }}%
\providecommand \urlprefix  [0]{URL }%
\providecommand \Eprint [0]{\href }%
\providecommand \doibase [0]{http://dx.doi.org/}%
\providecommand \selectlanguage [0]{\@gobble}%
\providecommand \bibinfo  [0]{\@secondoftwo}%
\providecommand \bibfield  [0]{\@secondoftwo}%
\providecommand \translation [1]{[#1]}%
\providecommand \BibitemOpen [0]{}%
\providecommand \bibitemStop [0]{}%
\providecommand \bibitemNoStop [0]{.\EOS\space}%
\providecommand \EOS [0]{\spacefactor3000\relax}%
\providecommand \BibitemShut  [1]{\csname bibitem#1\endcsname}%
\let\auto@bib@innerbib\@empty
\bibitem [{\citenamefont {Hinder}\ \emph {et~al.}(2018)\citenamefont {Hinder}, \citenamefont {Kidder},\ and\ \citenamefont {Pfeiffer}}]{Hinder:2017sxy}%
  \BibitemOpen
  \bibfield  {author} {\bibinfo {author} {\bibfnamefont {Ian}\ \bibnamefont {Hinder}}, \bibinfo {author} {\bibfnamefont {Lawrence~E.}\ \bibnamefont {Kidder}}, \ and\ \bibinfo {author} {\bibfnamefont {Harald~P.}\ \bibnamefont {Pfeiffer}},\ }\bibfield  {title} {\enquote {\bibinfo {title} {{Eccentric binary black hole inspiral-merger-ringdown gravitational waveform model from numerical relativity and post-Newtonian theory}},}\ }\href {\doibase 10.1103/PhysRevD.98.044015} {\bibfield  {journal} {\bibinfo  {journal} {Phys. Rev. D}\ }\textbf {\bibinfo {volume} {98}},\ \bibinfo {pages} {044015} (\bibinfo {year} {2018})},\ \Eprint {http://arxiv.org/abs/1709.02007} {arXiv:1709.02007 [gr-qc]} \BibitemShut {NoStop}%
\bibitem [{\citenamefont {Harry}(2010)}]{Harry:2010zz}%
  \BibitemOpen
  \bibfield  {author} {\bibinfo {author} {\bibfnamefont {Gregory~M.}\ \bibnamefont {Harry}} (\bibinfo {collaboration} {LIGO Scientific}),\ }\bibfield  {title} {\enquote {\bibinfo {title} {{Advanced LIGO: The next generation of gravitational wave detectors}},}\ }\href {\doibase 10.1088/0264-9381/27/8/084006} {\bibfield  {journal} {\bibinfo  {journal} {Class. Quant. Grav.}\ }\textbf {\bibinfo {volume} {27}},\ \bibinfo {pages} {084006} (\bibinfo {year} {2010})}\BibitemShut {NoStop}%
\bibitem [{\citenamefont {Acernese}\ \emph {et~al.}(2015)\citenamefont {Acernese} \emph {et~al.}}]{VIRGO:2014yos}%
  \BibitemOpen
  \bibfield  {author} {\bibinfo {author} {\bibfnamefont {F.}~\bibnamefont {Acernese}} \emph {et~al.} (\bibinfo {collaboration} {VIRGO}),\ }\bibfield  {title} {\enquote {\bibinfo {title} {{Advanced Virgo: a second-generation interferometric gravitational wave detector}},}\ }\href {\doibase 10.1088/0264-9381/32/2/024001} {\bibfield  {journal} {\bibinfo  {journal} {Class. Quant. Grav.}\ }\textbf {\bibinfo {volume} {32}},\ \bibinfo {pages} {024001} (\bibinfo {year} {2015})},\ \Eprint {http://arxiv.org/abs/1408.3978} {arXiv:1408.3978 [gr-qc]} \BibitemShut {NoStop}%
\bibitem [{\citenamefont {Akutsu}\ \emph {et~al.}(2021)\citenamefont {Akutsu} \emph {et~al.}}]{KAGRA:2020tym}%
  \BibitemOpen
  \bibfield  {author} {\bibinfo {author} {\bibfnamefont {T.}~\bibnamefont {Akutsu}} \emph {et~al.} (\bibinfo {collaboration} {KAGRA}),\ }\bibfield  {title} {\enquote {\bibinfo {title} {{Overview of KAGRA: Detector design and construction history}},}\ }\href {\doibase 10.1093/ptep/ptaa125} {\bibfield  {journal} {\bibinfo  {journal} {PTEP}\ }\textbf {\bibinfo {volume} {2021}},\ \bibinfo {pages} {05A101} (\bibinfo {year} {2021})},\ \Eprint {http://arxiv.org/abs/2005.05574} {arXiv:2005.05574 [physics.ins-det]} \BibitemShut {NoStop}%
\bibitem [{\citenamefont {Abbott}\ \emph {et~al.}(2019)\citenamefont {Abbott} \emph {et~al.}}]{LIGOScientific:2018mvr}%
  \BibitemOpen
  \bibfield  {author} {\bibinfo {author} {\bibfnamefont {B.~P.}\ \bibnamefont {Abbott}} \emph {et~al.} (\bibinfo {collaboration} {LIGO Scientific, Virgo}),\ }\bibfield  {title} {\enquote {\bibinfo {title} {{GWTC-1: A Gravitational-Wave Transient Catalog of Compact Binary Mergers Observed by LIGO and Virgo during the First and Second Observing Runs}},}\ }\href {\doibase 10.1103/PhysRevX.9.031040} {\bibfield  {journal} {\bibinfo  {journal} {Phys. Rev. X}\ }\textbf {\bibinfo {volume} {9}},\ \bibinfo {pages} {031040} (\bibinfo {year} {2019})},\ \Eprint {http://arxiv.org/abs/1811.12907} {arXiv:1811.12907 [astro-ph.HE]} \BibitemShut {NoStop}%
\bibitem [{\citenamefont {Abbott}\ \emph {et~al.}(2021{\natexlab{a}})\citenamefont {Abbott} \emph {et~al.}}]{LIGOScientific:2020ibl}%
  \BibitemOpen
  \bibfield  {author} {\bibinfo {author} {\bibfnamefont {R.}~\bibnamefont {Abbott}} \emph {et~al.} (\bibinfo {collaboration} {LIGO Scientific, Virgo}),\ }\bibfield  {title} {\enquote {\bibinfo {title} {{GWTC-2: Compact Binary Coalescences Observed by LIGO and Virgo During the First Half of the Third Observing Run}},}\ }\href {\doibase 10.1103/PhysRevX.11.021053} {\bibfield  {journal} {\bibinfo  {journal} {Phys. Rev. X}\ }\textbf {\bibinfo {volume} {11}},\ \bibinfo {pages} {021053} (\bibinfo {year} {2021}{\natexlab{a}})},\ \Eprint {http://arxiv.org/abs/2010.14527} {arXiv:2010.14527 [gr-qc]} \BibitemShut {NoStop}%
\bibitem [{\citenamefont {Abbott}\ \emph {et~al.}(2021{\natexlab{b}})\citenamefont {Abbott} \emph {et~al.}}]{LIGOScientific:2021usb}%
  \BibitemOpen
  \bibfield  {author} {\bibinfo {author} {\bibfnamefont {R.}~\bibnamefont {Abbott}} \emph {et~al.} (\bibinfo {collaboration} {LIGO Scientific, VIRGO}),\ }\bibfield  {title} {\enquote {\bibinfo {title} {{GWTC-2.1: Deep Extended Catalog of Compact Binary Coalescences Observed by LIGO and Virgo During the First Half of the Third Observing Run}},}\ }\href@noop {} {\  (\bibinfo {year} {2021}{\natexlab{b}})},\ \Eprint {http://arxiv.org/abs/2108.01045} {arXiv:2108.01045 [gr-qc]} \BibitemShut {NoStop}%
\bibitem [{\citenamefont {Abbott}\ \emph {et~al.}(2021{\natexlab{c}})\citenamefont {Abbott} \emph {et~al.}}]{LIGOScientific:2021djp}%
  \BibitemOpen
  \bibfield  {author} {\bibinfo {author} {\bibfnamefont {R.}~\bibnamefont {Abbott}} \emph {et~al.} (\bibinfo {collaboration} {LIGO Scientific, VIRGO, KAGRA}),\ }\bibfield  {title} {\enquote {\bibinfo {title} {{GWTC-3: Compact Binary Coalescences Observed by LIGO and Virgo During the Second Part of the Third Observing Run}},}\ }\href@noop {} {\  (\bibinfo {year} {2021}{\natexlab{c}})},\ \Eprint {http://arxiv.org/abs/2111.03606} {arXiv:2111.03606 [gr-qc]} \BibitemShut {NoStop}%
\bibitem [{\citenamefont {Gond\'an}\ and\ \citenamefont {Kocsis}(2021)}]{Gondan:2020svr}%
  \BibitemOpen
  \bibfield  {author} {\bibinfo {author} {\bibfnamefont {L\'aszl\'o}\ \bibnamefont {Gond\'an}}\ and\ \bibinfo {author} {\bibfnamefont {Bence}\ \bibnamefont {Kocsis}},\ }\bibfield  {title} {\enquote {\bibinfo {title} {{High eccentricities and high masses characterize gravitational-wave captures in galactic nuclei as seen by Earth-based detectors}},}\ }\href {\doibase 10.1093/mnras/stab1722} {\bibfield  {journal} {\bibinfo  {journal} {Mon. Not. Roy. Astron. Soc.}\ }\textbf {\bibinfo {volume} {506}},\ \bibinfo {pages} {1665--1696} (\bibinfo {year} {2021})},\ \Eprint {http://arxiv.org/abs/2011.02507} {arXiv:2011.02507 [astro-ph.HE]} \BibitemShut {NoStop}%
\bibitem [{\citenamefont {Romero-Shaw}\ \emph {et~al.}(2019)\citenamefont {Romero-Shaw}, \citenamefont {Lasky},\ and\ \citenamefont {Thrane}}]{Romero-Shaw:2019itr}%
  \BibitemOpen
  \bibfield  {author} {\bibinfo {author} {\bibfnamefont {Isobel~M.}\ \bibnamefont {Romero-Shaw}}, \bibinfo {author} {\bibfnamefont {Paul~D.}\ \bibnamefont {Lasky}}, \ and\ \bibinfo {author} {\bibfnamefont {Eric}\ \bibnamefont {Thrane}},\ }\bibfield  {title} {\enquote {\bibinfo {title} {{Searching for Eccentricity: Signatures of Dynamical Formation in the First Gravitational-Wave Transient Catalogue of LIGO and Virgo}},}\ }\href {\doibase 10.1093/mnras/stz2996} {\bibfield  {journal} {\bibinfo  {journal} {Mon. Not. Roy. Astron. Soc.}\ }\textbf {\bibinfo {volume} {490}},\ \bibinfo {pages} {5210--5216} (\bibinfo {year} {2019})},\ \Eprint {http://arxiv.org/abs/1909.05466} {arXiv:1909.05466 [astro-ph.HE]} \BibitemShut {NoStop}%
\bibitem [{\citenamefont {Abbott}\ \emph {et~al.}(2020)\citenamefont {Abbott} \emph {et~al.}}]{LIGOScientific:2020iuh}%
  \BibitemOpen
  \bibfield  {author} {\bibinfo {author} {\bibfnamefont {R.}~\bibnamefont {Abbott}} \emph {et~al.} (\bibinfo {collaboration} {LIGO Scientific, Virgo}),\ }\bibfield  {title} {\enquote {\bibinfo {title} {{GW190521: A Binary Black Hole Merger with a Total Mass of $150 M_{\odot}$}},}\ }\href {\doibase 10.1103/PhysRevLett.125.101102} {\bibfield  {journal} {\bibinfo  {journal} {Phys. Rev. Lett.}\ }\textbf {\bibinfo {volume} {125}},\ \bibinfo {pages} {101102} (\bibinfo {year} {2020})},\ \Eprint {http://arxiv.org/abs/2009.01075} {arXiv:2009.01075 [gr-qc]} \BibitemShut {NoStop}%
\bibitem [{\citenamefont {Romero-Shaw}\ \emph {et~al.}(2020)\citenamefont {Romero-Shaw}, \citenamefont {Lasky}, \citenamefont {Thrane},\ and\ \citenamefont {Bustillo}}]{RomeroShaw:2020thy}%
  \BibitemOpen
  \bibfield  {author} {\bibinfo {author} {\bibfnamefont {Isobel~M.}\ \bibnamefont {Romero-Shaw}}, \bibinfo {author} {\bibfnamefont {Paul~D.}\ \bibnamefont {Lasky}}, \bibinfo {author} {\bibfnamefont {Eric}\ \bibnamefont {Thrane}}, \ and\ \bibinfo {author} {\bibfnamefont {Juan~Calderon}\ \bibnamefont {Bustillo}},\ }\bibfield  {title} {\enquote {\bibinfo {title} {{GW190521: orbital eccentricity and signatures of dynamical formation in a binary black hole merger signal}},}\ }\href {\doibase 10.3847/2041-8213/abbe26} {\bibfield  {journal} {\bibinfo  {journal} {Astrophys. J. Lett.}\ }\textbf {\bibinfo {volume} {903}},\ \bibinfo {pages} {L5} (\bibinfo {year} {2020})},\ \Eprint {http://arxiv.org/abs/2009.04771} {arXiv:2009.04771 [astro-ph.HE]} \BibitemShut {NoStop}%
\bibitem [{\citenamefont {Gayathri}\ \emph {et~al.}(2022)\citenamefont {Gayathri}, \citenamefont {Healy}, \citenamefont {Lange}, \citenamefont {O'Brien}, \citenamefont {Szczepanczyk}, \citenamefont {Bartos}, \citenamefont {Campanelli}, \citenamefont {Klimenko}, \citenamefont {Lousto},\ and\ \citenamefont {O'Shaughnessy}}]{Gayathri:2020coq}%
  \BibitemOpen
  \bibfield  {author} {\bibinfo {author} {\bibfnamefont {V.}~\bibnamefont {Gayathri}}, \bibinfo {author} {\bibfnamefont {J.}~\bibnamefont {Healy}}, \bibinfo {author} {\bibfnamefont {J.}~\bibnamefont {Lange}}, \bibinfo {author} {\bibfnamefont {B.}~\bibnamefont {O'Brien}}, \bibinfo {author} {\bibfnamefont {M.}~\bibnamefont {Szczepanczyk}}, \bibinfo {author} {\bibfnamefont {Imre}\ \bibnamefont {Bartos}}, \bibinfo {author} {\bibfnamefont {M.}~\bibnamefont {Campanelli}}, \bibinfo {author} {\bibfnamefont {S.}~\bibnamefont {Klimenko}}, \bibinfo {author} {\bibfnamefont {C.~O.}\ \bibnamefont {Lousto}}, \ and\ \bibinfo {author} {\bibfnamefont {R.}~\bibnamefont {O'Shaughnessy}},\ }\bibfield  {title} {\enquote {\bibinfo {title} {{Eccentricity estimate for black hole mergers with numerical relativity simulations}},}\ }\href {\doibase 10.1038/s41550-021-01568-w} {\bibfield  {journal} {\bibinfo  {journal} {Nature Astron.}\ }\textbf {\bibinfo {volume} {6}},\ \bibinfo {pages} {344--349} (\bibinfo {year} {2022})},\ \Eprint
  {http://arxiv.org/abs/2009.05461} {arXiv:2009.05461 [astro-ph.HE]} \BibitemShut {NoStop}%
\bibitem [{\citenamefont {Calder\'on~Bustillo}\ \emph {et~al.}(2021)\citenamefont {Calder\'on~Bustillo}, \citenamefont {Sanchis-Gual}, \citenamefont {Torres-Forn\'e},\ and\ \citenamefont {Font}}]{CalderonBustillo:2020xms}%
  \BibitemOpen
  \bibfield  {author} {\bibinfo {author} {\bibfnamefont {Juan}\ \bibnamefont {Calder\'on~Bustillo}}, \bibinfo {author} {\bibfnamefont {Nicolas}\ \bibnamefont {Sanchis-Gual}}, \bibinfo {author} {\bibfnamefont {Alejandro}\ \bibnamefont {Torres-Forn\'e}}, \ and\ \bibinfo {author} {\bibfnamefont {Jos\'e~A.}\ \bibnamefont {Font}},\ }\bibfield  {title} {\enquote {\bibinfo {title} {{Confusing Head-On Collisions with Precessing Intermediate-Mass Binary Black Hole Mergers}},}\ }\href {\doibase 10.1103/PhysRevLett.126.201101} {\bibfield  {journal} {\bibinfo  {journal} {Phys. Rev. Lett.}\ }\textbf {\bibinfo {volume} {126}},\ \bibinfo {pages} {201101} (\bibinfo {year} {2021})},\ \Eprint {http://arxiv.org/abs/2009.01066} {arXiv:2009.01066 [gr-qc]} \BibitemShut {NoStop}%
\bibitem [{\citenamefont {Tiwari}\ \emph {et~al.}(2019)\citenamefont {Tiwari}, \citenamefont {Achamveedu}, \citenamefont {Haney},\ and\ \citenamefont {Hemantakumar}}]{Tiwari:2019jtz}%
  \BibitemOpen
  \bibfield  {author} {\bibinfo {author} {\bibfnamefont {Srishti}\ \bibnamefont {Tiwari}}, \bibinfo {author} {\bibfnamefont {Gopakumar}\ \bibnamefont {Achamveedu}}, \bibinfo {author} {\bibfnamefont {Maria}\ \bibnamefont {Haney}}, \ and\ \bibinfo {author} {\bibfnamefont {Phurailatapam}\ \bibnamefont {Hemantakumar}},\ }\bibfield  {title} {\enquote {\bibinfo {title} {{Ready-to-use Fourier domain templates for compact binaries inspiraling along moderately eccentric orbits}},}\ }\href {\doibase 10.1103/PhysRevD.99.124008} {\bibfield  {journal} {\bibinfo  {journal} {Phys. Rev. D}\ }\textbf {\bibinfo {volume} {99}},\ \bibinfo {pages} {124008} (\bibinfo {year} {2019})},\ \Eprint {http://arxiv.org/abs/1905.07956} {arXiv:1905.07956 [gr-qc]} \BibitemShut {NoStop}%
\bibitem [{\citenamefont {Huerta}\ \emph {et~al.}(2014)\citenamefont {Huerta}, \citenamefont {Kumar}, \citenamefont {McWilliams}, \citenamefont {O'Shaughnessy},\ and\ \citenamefont {Yunes}}]{Huerta:2014eca}%
  \BibitemOpen
  \bibfield  {author} {\bibinfo {author} {\bibfnamefont {E.~A.}\ \bibnamefont {Huerta}}, \bibinfo {author} {\bibfnamefont {Prayush}\ \bibnamefont {Kumar}}, \bibinfo {author} {\bibfnamefont {Sean~T.}\ \bibnamefont {McWilliams}}, \bibinfo {author} {\bibfnamefont {Richard}\ \bibnamefont {O'Shaughnessy}}, \ and\ \bibinfo {author} {\bibfnamefont {Nicol\'as}\ \bibnamefont {Yunes}},\ }\bibfield  {title} {\enquote {\bibinfo {title} {{Accurate and efficient waveforms for compact binaries on eccentric orbits}},}\ }\href {\doibase 10.1103/PhysRevD.90.084016} {\bibfield  {journal} {\bibinfo  {journal} {Phys. Rev. D}\ }\textbf {\bibinfo {volume} {90}},\ \bibinfo {pages} {084016} (\bibinfo {year} {2014})},\ \Eprint {http://arxiv.org/abs/1408.3406} {arXiv:1408.3406 [gr-qc]} \BibitemShut {NoStop}%
\bibitem [{\citenamefont {Moore}\ \emph {et~al.}(2016)\citenamefont {Moore}, \citenamefont {Favata}, \citenamefont {Arun},\ and\ \citenamefont {Mishra}}]{Moore:2016qxz}%
  \BibitemOpen
  \bibfield  {author} {\bibinfo {author} {\bibfnamefont {Blake}\ \bibnamefont {Moore}}, \bibinfo {author} {\bibfnamefont {Marc}\ \bibnamefont {Favata}}, \bibinfo {author} {\bibfnamefont {K.~G.}\ \bibnamefont {Arun}}, \ and\ \bibinfo {author} {\bibfnamefont {Chandra~Kant}\ \bibnamefont {Mishra}},\ }\bibfield  {title} {\enquote {\bibinfo {title} {{Gravitational-wave phasing for low-eccentricity inspiralling compact binaries to 3PN order}},}\ }\href {\doibase 10.1103/PhysRevD.93.124061} {\bibfield  {journal} {\bibinfo  {journal} {Phys. Rev. D}\ }\textbf {\bibinfo {volume} {93}},\ \bibinfo {pages} {124061} (\bibinfo {year} {2016})},\ \Eprint {http://arxiv.org/abs/1605.00304} {arXiv:1605.00304 [gr-qc]} \BibitemShut {NoStop}%
\bibitem [{\citenamefont {Damour}\ \emph {et~al.}(2004)\citenamefont {Damour}, \citenamefont {Gopakumar},\ and\ \citenamefont {Iyer}}]{Damour:2004bz}%
  \BibitemOpen
  \bibfield  {author} {\bibinfo {author} {\bibfnamefont {Thibault}\ \bibnamefont {Damour}}, \bibinfo {author} {\bibfnamefont {Achamveedu}\ \bibnamefont {Gopakumar}}, \ and\ \bibinfo {author} {\bibfnamefont {Bala~R.}\ \bibnamefont {Iyer}},\ }\bibfield  {title} {\enquote {\bibinfo {title} {{Phasing of gravitational waves from inspiralling eccentric binaries}},}\ }\href {\doibase 10.1103/PhysRevD.70.064028} {\bibfield  {journal} {\bibinfo  {journal} {Phys. Rev. D}\ }\textbf {\bibinfo {volume} {70}},\ \bibinfo {pages} {064028} (\bibinfo {year} {2004})},\ \Eprint {http://arxiv.org/abs/gr-qc/0404128} {arXiv:gr-qc/0404128} \BibitemShut {NoStop}%
\bibitem [{\citenamefont {Konigsdorffer}\ and\ \citenamefont {Gopakumar}(2006)}]{Konigsdorffer:2006zt}%
  \BibitemOpen
  \bibfield  {author} {\bibinfo {author} {\bibfnamefont {Christian}\ \bibnamefont {Konigsdorffer}}\ and\ \bibinfo {author} {\bibfnamefont {Achamveedu}\ \bibnamefont {Gopakumar}},\ }\bibfield  {title} {\enquote {\bibinfo {title} {{Phasing of gravitational waves from inspiralling eccentric binaries at the third-and-a-half post-Newtonian order}},}\ }\href {\doibase 10.1103/PhysRevD.73.124012} {\bibfield  {journal} {\bibinfo  {journal} {Phys. Rev. D}\ }\textbf {\bibinfo {volume} {73}},\ \bibinfo {pages} {124012} (\bibinfo {year} {2006})},\ \Eprint {http://arxiv.org/abs/gr-qc/0603056} {arXiv:gr-qc/0603056} \BibitemShut {NoStop}%
\bibitem [{\citenamefont {Memmesheimer}\ \emph {et~al.}(2004)\citenamefont {Memmesheimer}, \citenamefont {Gopakumar},\ and\ \citenamefont {Schaefer}}]{Memmesheimer:2004cv}%
  \BibitemOpen
  \bibfield  {author} {\bibinfo {author} {\bibfnamefont {Raoul-Martin}\ \bibnamefont {Memmesheimer}}, \bibinfo {author} {\bibfnamefont {Achamveedu}\ \bibnamefont {Gopakumar}}, \ and\ \bibinfo {author} {\bibfnamefont {Gerhard}\ \bibnamefont {Schaefer}},\ }\bibfield  {title} {\enquote {\bibinfo {title} {{Third post-Newtonian accurate generalized quasi-Keplerian parametrization for compact binaries in eccentric orbits}},}\ }\href {\doibase 10.1103/PhysRevD.70.104011} {\bibfield  {journal} {\bibinfo  {journal} {Phys. Rev. D}\ }\textbf {\bibinfo {volume} {70}},\ \bibinfo {pages} {104011} (\bibinfo {year} {2004})},\ \Eprint {http://arxiv.org/abs/gr-qc/0407049} {arXiv:gr-qc/0407049} \BibitemShut {NoStop}%
\bibitem [{\citenamefont {Cho}\ \emph {et~al.}(2022)\citenamefont {Cho}, \citenamefont {Tanay}, \citenamefont {Gopakumar},\ and\ \citenamefont {Lee}}]{Cho:2021oai}%
  \BibitemOpen
  \bibfield  {author} {\bibinfo {author} {\bibfnamefont {Gihyuk}\ \bibnamefont {Cho}}, \bibinfo {author} {\bibfnamefont {Sashwat}\ \bibnamefont {Tanay}}, \bibinfo {author} {\bibfnamefont {Achamveedu}\ \bibnamefont {Gopakumar}}, \ and\ \bibinfo {author} {\bibfnamefont {Hyung~Mok}\ \bibnamefont {Lee}},\ }\bibfield  {title} {\enquote {\bibinfo {title} {{Generalized quasi-Keplerian solution for eccentric, nonspinning compact binaries at 4PN order and the associated inspiral-merger-ringdown waveform}},}\ }\href {\doibase 10.1103/PhysRevD.105.064010} {\bibfield  {journal} {\bibinfo  {journal} {Phys. Rev. D}\ }\textbf {\bibinfo {volume} {105}},\ \bibinfo {pages} {064010} (\bibinfo {year} {2022})},\ \Eprint {http://arxiv.org/abs/2110.09608} {arXiv:2110.09608 [gr-qc]} \BibitemShut {NoStop}%
\bibitem [{\citenamefont {Chattaraj}\ \emph {et~al.}(2022)\citenamefont {Chattaraj}, \citenamefont {RoyChowdhury}, \citenamefont {Divyajyoti}, \citenamefont {Mishra},\ and\ \citenamefont {Gupta}}]{Chattaraj:2022tay}%
  \BibitemOpen
  \bibfield  {author} {\bibinfo {author} {\bibfnamefont {Abhishek}\ \bibnamefont {Chattaraj}}, \bibinfo {author} {\bibfnamefont {Tamal}\ \bibnamefont {RoyChowdhury}}, \bibinfo {author} {\bibnamefont {Divyajyoti}}, \bibinfo {author} {\bibfnamefont {Chandra~Kant}\ \bibnamefont {Mishra}}, \ and\ \bibinfo {author} {\bibfnamefont {Anshu}\ \bibnamefont {Gupta}},\ }\bibfield  {title} {\enquote {\bibinfo {title} {{High accuracy post-Newtonian and numerical relativity comparisons involving higher modes for eccentric binary black holes and a dominant mode eccentric inspiral-merger-ringdown model}},}\ }\href {\doibase 10.1103/PhysRevD.106.124008} {\bibfield  {journal} {\bibinfo  {journal} {Phys. Rev. D}\ }\textbf {\bibinfo {volume} {106}},\ \bibinfo {pages} {124008} (\bibinfo {year} {2022})},\ \Eprint {http://arxiv.org/abs/2204.02377} {arXiv:2204.02377 [gr-qc]} \BibitemShut {NoStop}%
\bibitem [{\citenamefont {Hinderer}\ and\ \citenamefont {Babak}(2017)}]{Hinderer:2017jcs}%
  \BibitemOpen
  \bibfield  {author} {\bibinfo {author} {\bibfnamefont {Tanja}\ \bibnamefont {Hinderer}}\ and\ \bibinfo {author} {\bibfnamefont {Stanislav}\ \bibnamefont {Babak}},\ }\bibfield  {title} {\enquote {\bibinfo {title} {{Foundations of an effective-one-body model for coalescing binaries on eccentric orbits}},}\ }\href {\doibase 10.1103/PhysRevD.96.104048} {\bibfield  {journal} {\bibinfo  {journal} {Phys. Rev. D}\ }\textbf {\bibinfo {volume} {96}},\ \bibinfo {pages} {104048} (\bibinfo {year} {2017})},\ \Eprint {http://arxiv.org/abs/1707.08426} {arXiv:1707.08426 [gr-qc]} \BibitemShut {NoStop}%
\bibitem [{\citenamefont {Cao}\ and\ \citenamefont {Han}(2017)}]{Cao:2017ndf}%
  \BibitemOpen
  \bibfield  {author} {\bibinfo {author} {\bibfnamefont {Zhoujian}\ \bibnamefont {Cao}}\ and\ \bibinfo {author} {\bibfnamefont {Wen-Biao}\ \bibnamefont {Han}},\ }\bibfield  {title} {\enquote {\bibinfo {title} {{Waveform model for an eccentric binary black hole based on the effective-one-body-numerical-relativity formalism}},}\ }\href {\doibase 10.1103/PhysRevD.96.044028} {\bibfield  {journal} {\bibinfo  {journal} {Phys. Rev. D}\ }\textbf {\bibinfo {volume} {96}},\ \bibinfo {pages} {044028} (\bibinfo {year} {2017})},\ \Eprint {http://arxiv.org/abs/1708.00166} {arXiv:1708.00166 [gr-qc]} \BibitemShut {NoStop}%
\bibitem [{\citenamefont {Chiaramello}\ and\ \citenamefont {Nagar}(2020)}]{Chiaramello:2020ehz}%
  \BibitemOpen
  \bibfield  {author} {\bibinfo {author} {\bibfnamefont {Danilo}\ \bibnamefont {Chiaramello}}\ and\ \bibinfo {author} {\bibfnamefont {Alessandro}\ \bibnamefont {Nagar}},\ }\bibfield  {title} {\enquote {\bibinfo {title} {{Faithful analytical effective-one-body waveform model for spin-aligned, moderately eccentric, coalescing black hole binaries}},}\ }\href {\doibase 10.1103/PhysRevD.101.101501} {\bibfield  {journal} {\bibinfo  {journal} {Phys. Rev. D}\ }\textbf {\bibinfo {volume} {101}},\ \bibinfo {pages} {101501} (\bibinfo {year} {2020})},\ \Eprint {http://arxiv.org/abs/2001.11736} {arXiv:2001.11736 [gr-qc]} \BibitemShut {NoStop}%
\bibitem [{\citenamefont {Albanesi}\ \emph {et~al.}(2023)\citenamefont {Albanesi}, \citenamefont {Bernuzzi}, \citenamefont {Damour}, \citenamefont {Nagar},\ and\ \citenamefont {Placidi}}]{Albanesi:2023bgi}%
  \BibitemOpen
  \bibfield  {author} {\bibinfo {author} {\bibfnamefont {Simone}\ \bibnamefont {Albanesi}}, \bibinfo {author} {\bibfnamefont {Sebastiano}\ \bibnamefont {Bernuzzi}}, \bibinfo {author} {\bibfnamefont {Thibault}\ \bibnamefont {Damour}}, \bibinfo {author} {\bibfnamefont {Alessandro}\ \bibnamefont {Nagar}}, \ and\ \bibinfo {author} {\bibfnamefont {Andrea}\ \bibnamefont {Placidi}},\ }\bibfield  {title} {\enquote {\bibinfo {title} {{Faithful effective-one-body waveform of small-mass-ratio coalescing black hole binaries: The eccentric, nonspinning case}},}\ }\href {\doibase 10.1103/PhysRevD.108.084037} {\bibfield  {journal} {\bibinfo  {journal} {Phys. Rev. D}\ }\textbf {\bibinfo {volume} {108}},\ \bibinfo {pages} {084037} (\bibinfo {year} {2023})},\ \Eprint {http://arxiv.org/abs/2305.19336} {arXiv:2305.19336 [gr-qc]} \BibitemShut {NoStop}%
\bibitem [{\citenamefont {Albanesi}\ \emph {et~al.}(2022)\citenamefont {Albanesi}, \citenamefont {Placidi}, \citenamefont {Nagar}, \citenamefont {Orselli},\ and\ \citenamefont {Bernuzzi}}]{Albanesi:2022xge}%
  \BibitemOpen
  \bibfield  {author} {\bibinfo {author} {\bibfnamefont {Simone}\ \bibnamefont {Albanesi}}, \bibinfo {author} {\bibfnamefont {Andrea}\ \bibnamefont {Placidi}}, \bibinfo {author} {\bibfnamefont {Alessandro}\ \bibnamefont {Nagar}}, \bibinfo {author} {\bibfnamefont {Marta}\ \bibnamefont {Orselli}}, \ and\ \bibinfo {author} {\bibfnamefont {Sebastiano}\ \bibnamefont {Bernuzzi}},\ }\bibfield  {title} {\enquote {\bibinfo {title} {{New avenue for accurate analytical waveforms and fluxes for eccentric compact binaries}},}\ }\href {\doibase 10.1103/PhysRevD.105.L121503} {\bibfield  {journal} {\bibinfo  {journal} {Phys. Rev. D}\ }\textbf {\bibinfo {volume} {105}},\ \bibinfo {pages} {L121503} (\bibinfo {year} {2022})},\ \Eprint {http://arxiv.org/abs/2203.16286} {arXiv:2203.16286 [gr-qc]} \BibitemShut {NoStop}%
\bibitem [{\citenamefont {Riemenschneider}\ \emph {et~al.}(2021)\citenamefont {Riemenschneider}, \citenamefont {Rettegno}, \citenamefont {Breschi}, \citenamefont {Albertini}, \citenamefont {Gamba}, \citenamefont {Bernuzzi},\ and\ \citenamefont {Nagar}}]{Riemenschneider:2021ppj}%
  \BibitemOpen
  \bibfield  {author} {\bibinfo {author} {\bibfnamefont {Gunnar}\ \bibnamefont {Riemenschneider}}, \bibinfo {author} {\bibfnamefont {Piero}\ \bibnamefont {Rettegno}}, \bibinfo {author} {\bibfnamefont {Matteo}\ \bibnamefont {Breschi}}, \bibinfo {author} {\bibfnamefont {Angelica}\ \bibnamefont {Albertini}}, \bibinfo {author} {\bibfnamefont {Rossella}\ \bibnamefont {Gamba}}, \bibinfo {author} {\bibfnamefont {Sebastiano}\ \bibnamefont {Bernuzzi}}, \ and\ \bibinfo {author} {\bibfnamefont {Alessandro}\ \bibnamefont {Nagar}},\ }\bibfield  {title} {\enquote {\bibinfo {title} {{Assessment of consistent next-to-quasicircular corrections and postadiabatic approximation in effective-one-body multipolar waveforms for binary black hole coalescences}},}\ }\href {\doibase 10.1103/PhysRevD.104.104045} {\bibfield  {journal} {\bibinfo  {journal} {Phys. Rev. D}\ }\textbf {\bibinfo {volume} {104}},\ \bibinfo {pages} {104045} (\bibinfo {year} {2021})},\ \Eprint {http://arxiv.org/abs/2104.07533} {arXiv:2104.07533 [gr-qc]}
  \BibitemShut {NoStop}%
\bibitem [{\citenamefont {Ramos-Buades}\ \emph {et~al.}(2022)\citenamefont {Ramos-Buades}, \citenamefont {Buonanno}, \citenamefont {Khalil},\ and\ \citenamefont {Ossokine}}]{Ramos-Buades:2021adz}%
  \BibitemOpen
  \bibfield  {author} {\bibinfo {author} {\bibfnamefont {Antoni}\ \bibnamefont {Ramos-Buades}}, \bibinfo {author} {\bibfnamefont {Alessandra}\ \bibnamefont {Buonanno}}, \bibinfo {author} {\bibfnamefont {Mohammed}\ \bibnamefont {Khalil}}, \ and\ \bibinfo {author} {\bibfnamefont {Serguei}\ \bibnamefont {Ossokine}},\ }\bibfield  {title} {\enquote {\bibinfo {title} {{Effective-one-body multipolar waveforms for eccentric binary black holes with nonprecessing spins}},}\ }\href {\doibase 10.1103/PhysRevD.105.044035} {\bibfield  {journal} {\bibinfo  {journal} {Phys. Rev. D}\ }\textbf {\bibinfo {volume} {105}},\ \bibinfo {pages} {044035} (\bibinfo {year} {2022})},\ \Eprint {http://arxiv.org/abs/2112.06952} {arXiv:2112.06952 [gr-qc]} \BibitemShut {NoStop}%
\bibitem [{\citenamefont {Liu}\ \emph {et~al.}(2023)\citenamefont {Liu}, \citenamefont {Cao},\ and\ \citenamefont {Zhu}}]{Liu:2023ldr}%
  \BibitemOpen
  \bibfield  {author} {\bibinfo {author} {\bibfnamefont {Xiaolin}\ \bibnamefont {Liu}}, \bibinfo {author} {\bibfnamefont {Zhoujian}\ \bibnamefont {Cao}}, \ and\ \bibinfo {author} {\bibfnamefont {Zong-Hong}\ \bibnamefont {Zhu}},\ }\bibfield  {title} {\enquote {\bibinfo {title} {{Effective-One-Body Numerical-Relativity waveform model for Eccentric spin-precessing binary black hole coalescence}},}\ }\href@noop {} {\  (\bibinfo {year} {2023})},\ \Eprint {http://arxiv.org/abs/2310.04552} {arXiv:2310.04552 [gr-qc]} \BibitemShut {NoStop}%
\bibitem [{\citenamefont {Huerta}\ \emph {et~al.}(2017)\citenamefont {Huerta} \emph {et~al.}}]{Huerta:2016rwp}%
  \BibitemOpen
  \bibfield  {author} {\bibinfo {author} {\bibfnamefont {E.~A.}\ \bibnamefont {Huerta}} \emph {et~al.},\ }\bibfield  {title} {\enquote {\bibinfo {title} {{Complete waveform model for compact binaries on eccentric orbits}},}\ }\href {\doibase 10.1103/PhysRevD.95.024038} {\bibfield  {journal} {\bibinfo  {journal} {Phys. Rev. D}\ }\textbf {\bibinfo {volume} {95}},\ \bibinfo {pages} {024038} (\bibinfo {year} {2017})},\ \Eprint {http://arxiv.org/abs/1609.05933} {arXiv:1609.05933 [gr-qc]} \BibitemShut {NoStop}%
\bibitem [{\citenamefont {Huerta}\ \emph {et~al.}(2018)\citenamefont {Huerta} \emph {et~al.}}]{Huerta:2017kez}%
  \BibitemOpen
  \bibfield  {author} {\bibinfo {author} {\bibfnamefont {E.~A.}\ \bibnamefont {Huerta}} \emph {et~al.},\ }\bibfield  {title} {\enquote {\bibinfo {title} {{Eccentric, nonspinning, inspiral, Gaussian-process merger approximant for the detection and characterization of eccentric binary black hole mergers}},}\ }\href {\doibase 10.1103/PhysRevD.97.024031} {\bibfield  {journal} {\bibinfo  {journal} {Phys. Rev. D}\ }\textbf {\bibinfo {volume} {97}},\ \bibinfo {pages} {024031} (\bibinfo {year} {2018})},\ \Eprint {http://arxiv.org/abs/1711.06276} {arXiv:1711.06276 [gr-qc]} \BibitemShut {NoStop}%
\bibitem [{\citenamefont {Joshi}\ \emph {et~al.}(2023)\citenamefont {Joshi}, \citenamefont {Rosofsky}, \citenamefont {Haas},\ and\ \citenamefont {Huerta}}]{Joshi:2022ocr}%
  \BibitemOpen
  \bibfield  {author} {\bibinfo {author} {\bibfnamefont {Abhishek~V.}\ \bibnamefont {Joshi}}, \bibinfo {author} {\bibfnamefont {Shawn~G.}\ \bibnamefont {Rosofsky}}, \bibinfo {author} {\bibfnamefont {Roland}\ \bibnamefont {Haas}}, \ and\ \bibinfo {author} {\bibfnamefont {E.~A.}\ \bibnamefont {Huerta}},\ }\bibfield  {title} {\enquote {\bibinfo {title} {{Numerical relativity higher order gravitational waveforms of eccentric, spinning, nonprecessing binary black hole mergers}},}\ }\href {\doibase 10.1103/PhysRevD.107.064038} {\bibfield  {journal} {\bibinfo  {journal} {Phys. Rev. D}\ }\textbf {\bibinfo {volume} {107}},\ \bibinfo {pages} {064038} (\bibinfo {year} {2023})},\ \Eprint {http://arxiv.org/abs/2210.01852} {arXiv:2210.01852 [gr-qc]} \BibitemShut {NoStop}%
\bibitem [{\citenamefont {Setyawati}\ and\ \citenamefont {Ohme}(2021)}]{Setyawati:2021gom}%
  \BibitemOpen
  \bibfield  {author} {\bibinfo {author} {\bibfnamefont {Yoshinta}\ \bibnamefont {Setyawati}}\ and\ \bibinfo {author} {\bibfnamefont {Frank}\ \bibnamefont {Ohme}},\ }\bibfield  {title} {\enquote {\bibinfo {title} {{Adding eccentricity to quasicircular binary-black-hole waveform models}},}\ }\href {\doibase 10.1103/PhysRevD.103.124011} {\bibfield  {journal} {\bibinfo  {journal} {Phys. Rev. D}\ }\textbf {\bibinfo {volume} {103}},\ \bibinfo {pages} {124011} (\bibinfo {year} {2021})},\ \Eprint {http://arxiv.org/abs/2101.11033} {arXiv:2101.11033 [gr-qc]} \BibitemShut {NoStop}%
\bibitem [{\citenamefont {Wang}\ \emph {et~al.}(2023)\citenamefont {Wang}, \citenamefont {Zou},\ and\ \citenamefont {Liu}}]{Wang:2023ueg}%
  \BibitemOpen
  \bibfield  {author} {\bibinfo {author} {\bibfnamefont {Hao}\ \bibnamefont {Wang}}, \bibinfo {author} {\bibfnamefont {Yuan-Chuan}\ \bibnamefont {Zou}}, \ and\ \bibinfo {author} {\bibfnamefont {Yu}~\bibnamefont {Liu}},\ }\bibfield  {title} {\enquote {\bibinfo {title} {{Phenomenological relationship between eccentric and quasicircular orbital binary black hole waveform}},}\ }\href {\doibase 10.1103/PhysRevD.107.124061} {\bibfield  {journal} {\bibinfo  {journal} {Phys. Rev. D}\ }\textbf {\bibinfo {volume} {107}},\ \bibinfo {pages} {124061} (\bibinfo {year} {2023})},\ \Eprint {http://arxiv.org/abs/2302.11227} {arXiv:2302.11227 [gr-qc]} \BibitemShut {NoStop}%
\bibitem [{\citenamefont {Islam}\ \emph {et~al.}(2021)\citenamefont {Islam}, \citenamefont {Varma}, \citenamefont {Lodman}, \citenamefont {Field}, \citenamefont {Khanna}, \citenamefont {Scheel}, \citenamefont {Pfeiffer}, \citenamefont {Gerosa},\ and\ \citenamefont {Kidder}}]{Islam:2021mha}%
  \BibitemOpen
  \bibfield  {author} {\bibinfo {author} {\bibfnamefont {Tousif}\ \bibnamefont {Islam}}, \bibinfo {author} {\bibfnamefont {Vijay}\ \bibnamefont {Varma}}, \bibinfo {author} {\bibfnamefont {Jackie}\ \bibnamefont {Lodman}}, \bibinfo {author} {\bibfnamefont {Scott~E.}\ \bibnamefont {Field}}, \bibinfo {author} {\bibfnamefont {Gaurav}\ \bibnamefont {Khanna}}, \bibinfo {author} {\bibfnamefont {Mark~A.}\ \bibnamefont {Scheel}}, \bibinfo {author} {\bibfnamefont {Harald~P.}\ \bibnamefont {Pfeiffer}}, \bibinfo {author} {\bibfnamefont {Davide}\ \bibnamefont {Gerosa}}, \ and\ \bibinfo {author} {\bibfnamefont {Lawrence~E.}\ \bibnamefont {Kidder}},\ }\bibfield  {title} {\enquote {\bibinfo {title} {{Eccentric binary black hole surrogate models for the gravitational waveform and remnant properties: comparable mass, nonspinning case}},}\ }\href {\doibase 10.1103/PhysRevD.103.064022} {\bibfield  {journal} {\bibinfo  {journal} {Phys. Rev. D}\ }\textbf {\bibinfo {volume} {103}},\ \bibinfo {pages} {064022} (\bibinfo {year}
  {2021})},\ \Eprint {http://arxiv.org/abs/2101.11798} {arXiv:2101.11798 [gr-qc]} \BibitemShut {NoStop}%
\bibitem [{\citenamefont {Hinder}\ \emph {et~al.}(2010)\citenamefont {Hinder}, \citenamefont {Herrmann}, \citenamefont {Laguna},\ and\ \citenamefont {Shoemaker}}]{Hinder:2008kv}%
  \BibitemOpen
  \bibfield  {author} {\bibinfo {author} {\bibfnamefont {Ian}\ \bibnamefont {Hinder}}, \bibinfo {author} {\bibfnamefont {Frank}\ \bibnamefont {Herrmann}}, \bibinfo {author} {\bibfnamefont {Pablo}\ \bibnamefont {Laguna}}, \ and\ \bibinfo {author} {\bibfnamefont {Deirdre}\ \bibnamefont {Shoemaker}},\ }\bibfield  {title} {\enquote {\bibinfo {title} {{Comparisons of eccentric binary black hole simulations with post-Newtonian models}},}\ }\href {\doibase 10.1103/PhysRevD.82.024033} {\bibfield  {journal} {\bibinfo  {journal} {Phys. Rev. D}\ }\textbf {\bibinfo {volume} {82}},\ \bibinfo {pages} {024033} (\bibinfo {year} {2010})},\ \Eprint {http://arxiv.org/abs/0806.1037} {arXiv:0806.1037 [gr-qc]} \BibitemShut {NoStop}%
\bibitem [{\citenamefont {Mora}\ and\ \citenamefont {Will}(2002)}]{Mora:2002gf}%
  \BibitemOpen
  \bibfield  {author} {\bibinfo {author} {\bibfnamefont {Thierry}\ \bibnamefont {Mora}}\ and\ \bibinfo {author} {\bibfnamefont {Clifford~M.}\ \bibnamefont {Will}},\ }\bibfield  {title} {\enquote {\bibinfo {title} {{Numerically generated quasiequilibrium orbits of black holes: Circular or eccentric?}}}\ }\href {\doibase 10.1103/PhysRevD.66.101501} {\bibfield  {journal} {\bibinfo  {journal} {Phys. Rev. D}\ }\textbf {\bibinfo {volume} {66}},\ \bibinfo {pages} {101501} (\bibinfo {year} {2002})},\ \Eprint {http://arxiv.org/abs/gr-qc/0208089} {arXiv:gr-qc/0208089} \BibitemShut {NoStop}%
\bibitem [{\citenamefont {McKechan}\ \emph {et~al.}(2010)\citenamefont {McKechan}, \citenamefont {Robinson},\ and\ \citenamefont {Sathyaprakash}}]{McKechan:2010kp}%
  \BibitemOpen
  \bibfield  {author} {\bibinfo {author} {\bibfnamefont {D.~J.~A.}\ \bibnamefont {McKechan}}, \bibinfo {author} {\bibfnamefont {C.}~\bibnamefont {Robinson}}, \ and\ \bibinfo {author} {\bibfnamefont {B.~S.}\ \bibnamefont {Sathyaprakash}},\ }\bibfield  {title} {\enquote {\bibinfo {title} {{A tapering window for time-domain templates and simulated signals in the detection of gravitational waves from coalescing compact binaries}},}\ }\bibfield  {booktitle} {\emph {\bibinfo {booktitle} {{Gravitational waves. Proceedings, 8th Edoardo Amaldi Conference, Amaldi 8, New York, USA, June 22-26, 2009}}},\ }\href {\doibase 10.1088/0264-9381/27/8/084020} {\bibfield  {journal} {\bibinfo  {journal} {Class. Quant. Grav.}\ }\textbf {\bibinfo {volume} {27}},\ \bibinfo {pages} {084020} (\bibinfo {year} {2010})},\ \Eprint {http://arxiv.org/abs/1003.2939} {arXiv:1003.2939 [gr-qc]} \BibitemShut {NoStop}%
\end{thebibliography}%


\end{document}